\documentclass[prd,aps,showpacs,nofootinbib,preprintnumbers,twocolumn]{revtex4-1}
\usepackage{graphicx,graphics,epsfig,epstopdf,subfigure,color,psfrag}
\usepackage{hyperref}
\newcommand{\be}{\begin{equation}}
\newcommand{\ee}{\end{equation}}
\newcommand{\bea}{\begin{eqnarray}}
\newcommand{\eea}{\end{eqnarray}}
\newcommand{\nn}{\nonumber}
\newcommand{\dd}{\displaystyle}

\newcommand{\ket}[1]{\left| #1 \right\rangle}

\begin{document}

\preprint {BARI-TH/2014-690}

\title{Exclusive  $b \to s  \nu \bar \nu$ induced  transitions  in RS$_c$ model}
\author{ P.~Biancofiore$^{a,b}$, P.~Colangelo$^{b}$,  F.~De~Fazio$^{b}$, E.~Scrimieri$^{a,b}$}
\affiliation{
$^{a}$Dipartimento Interateneo di Fisica, via Orabona 4, I-70126 Bari, Italy\\
$^{b}$INFN, Sezione di Bari, via Orabona 4, I-70126 Bari, Italy}
\begin{abstract}
We study a set of exclusive $B$ and $B_s$ decay modes induced by the rare $b \to s \nu \bar \nu$ transition in the RS$_c$ model, an extra-dimensional extension of the standard model
with warped 5D metric and  extended gauge group. We emphasize the role of correlations among the observables, and their  importance for detecting the predicted small deviations from the standard model expectations.
\end{abstract}
\pacs{1260.Cn, 1260.Fr, 1320.He}
\maketitle

\section{Introduction}
The current  searches for deviations from (or for further confirmation of)  the Standard Model  (SM) involve observables of increasing sophistication and difficulty. This is what happens for several quark flavour observables that are able to provide us with access to large energy scales,  complementing the  direct searches  at the CERN LHC \cite{general-review}.
The flavour changing neutral current (FCNC) processes,  loop-induced  and heavily suppressed in SM,   play a prominent role, and  an important  case to be studied is
the  $b \to s \nu \bar \nu$ transition,  which  in the Standard Model  proceeds through $Z^0$ penguin and box diagrams dominated by the contribution with the intermediate top quark \cite{Buras:1998raa}.

Rare $b$ decays with neutrino pairs in the final state are  experimentally challenging. Nevertheless,  the advent of new high-luminosity $B$ factories opens the possibility to access these modes
which, on the other hand, present remarkable features of theoretical clearness, as we discuss below.
We are mainly interested in the exclusive $B \to K \nu \bar \nu$  and $B \to K^* \nu \bar \nu$ decays, the branching fractions of which were predicted in SM of  ${\cal O}(10^{-6})$ \cite{Colangelo:1996ay,Melikhov:1998ug}.  Since the  results are affected by the  uncertainty  of the   form factors  parametrizing the hadronic matrix elements, particular attention has to be paid to such an issue.
Using form factors  from light-cone QCD sum rules together with experimental information on the $B \to K^* \gamma$ decay rate  \cite{Altmannshofer:2008dz},  new    predictions were  obtained in  SM \cite{Altmannshofer:2009ma},
\bea
{\cal B}(B^+ \to K^+ \nu \bar \nu) &=& (4.5 \pm 0.7) \times 10^{-6} \nn \\
{\cal B}(B \to K^* \nu \bar \nu) &=& (6.8 \pm^{1.0}_{1.1}) \times 10^{-6} \,\,
\label{LCSR}
\eea
(considering in the final state the sum over the three neutrino species),  that must be compared to the present experimental upper bounds.
 The  Belle Collaboration has established the limits, at 90$\%$ C.L.  \cite{Lutz:2013ftz},
\bea
{\cal B}(B^+ \to K^+ \nu \bar \nu) &<& 5.5\times 10^{-5} \nn \\
{\cal B}(B^0 \to K^0_S \nu \bar \nu) &<& 9.7\times 10^{-5} \nn \\
{\cal B}(B^+ \to K^{*+} \nu \bar \nu) &<& 4.0\times 10^{-5} \label{belle} \\
{\cal B}(B^0 \to K^{*0} \nu \bar \nu) &<& 5.5\times 10^{-5} \nn  \,\,.\eea
The bounds (at 90$\%$ C.L.) obtained by the BaBar Collaboration \cite{Lees:2013kla},
\bea
{\cal B}(B^+ \to K^+ \nu \bar \nu) &<& 1.6\times 10^{-5} \nn \\
{\cal B}(B^0 \to K^0 \nu \bar \nu) &<& 4.9\times 10^{-5} \nn \\
{\cal B}(B^+ \to K^{*+} \nu \bar \nu) &<& 6.4\times 10^{-5} \label{Babar}  \\
{\cal B}(B^0 \to K^{*0} \nu \bar \nu) &<& 12\times 10^{-5}  \nn \,\,, \eea
 are derived  combining  the results of the semileptonic tag reconstruction method \cite{delAmoSanchez:2010bk}  and of the hadronic tag reconstruction method \cite{Lees:2013kla}.

In addition to $B \to K^{(*)} \nu \bar \nu$, other  modes are induced by the $b \to s \nu \bar \nu$ transition, namely $B_s \to (\phi, \eta, \eta^\prime, f_0(980)) \nu \bar \nu$ that we also discuss in the following.  At present, the  experimental upper bounds for their rates are still quite high  \cite{Beringer:1900zz,Amhis:2012bh}, however they are also expected to be sizeably reduced at  the new  high-luminosity $B$ facilities.

The importance of the rare $b \to s \nu \bar \nu$  process relies on its  particular sensitivity  to new interactions.
In \cite{Kim:1999waa}  the effects of scalar and  tensor interactions have been discussed, with  particular attention to the distortion of the $q^2$ spectra (with $q^2$  the dilepton squared  four-momentum) with respect to  SM. The role of new right-handed operators has also been discussed  \cite{Melikhov:1998ug}, and
 the possibility of non-standard $Z$ couplings to $b$ and $s$ quarks has been considered   \cite{Buchalla:2000sk}.
An overview of the effects predicted in several new physics (NP) scenarios is in Ref. \cite{Altmannshofer:2009ma}.
In an analysis of the effects of a new neutral gauge boson $Z^\prime$,  the correlations between the branching ratios, as well as between these modes and the decay $B_s \to \mu^+ \mu^-$, have been  analyzed under different assumptions for  the $Z^\prime$ couplings  \cite{Buras:2012jb}.
In  extensions of SM based on  additional spatial dimensions,
predictions have been given  for the decay rates and distributions in minimal models with a single universal extra-dimension  \cite{Colangelo:2006vm}. Here, we  consider  the case of a single warped extra-dimension, as formalized in the Randall-Sundrum model \cite{RandallSundrum}, in particular in the realization with custodial protection of the $Z b_L \bar b_L$ coupling \cite{contino,carena,cacciapaglia}.
In \cite{buras2}  a range for the $B \to K^{(*)} \nu \bar \nu$ branching fractions has been predicted in this framework.
Here we  extend the analysis focusing on  other observables,  such as  several  differential distributions,  and on  various correlations,  reconsidering  the predictions   using  model parameters  singled out in a study of  the rare semileptonic $B \to K^* \ell^+ \ell^-$  modes \cite{Biancofiore:2014wpa}.

 In section \ref{sec:hamil} we describe the general form of the effective $b \to s  \nu \bar \nu$ Hamiltonian, and in sect. \ref{Kmode}
we define several  $B \to K^{(*)}  \nu \bar \nu$ observables. Generalities of the custodially-protected Randall-Sundrum model are described in sects.~\ref{sect:RS} and \ref{sect:WRS}, with particular attention to the parameter space bound for the model.
The  predictions are presented in sects.~\ref{sect:BKstar}, \ref{sect:RHcurrents} and \ref{sect:Bs}, with a discussion of  possible improvements. The conclusions are collected in the last section.

\section{$b \to s \nu \bar \nu$ effective hamiltonian }\label{sec:hamil}
In  SM the effective $b \to s \nu \bar \nu$ Hamiltonian   is written as
\bea
H_{eff}^{SM}&=&  {G_F \over \sqrt{2}} {\alpha \over 2 \pi \sin^2\theta_W} V_{tb}^* V_{ts} X(x_t) ({\bar b} s)_{V-A}({\bar \nu} \nu)_{V-A} \nn \\
&\equiv&C_L^{SM} O_L \,\,\ ,
\label{heff}
\eea
with   $O_L=({\bar b} s)_{V-A}({\bar \nu} \nu)_{V-A}$ \cite{Buras:1998raa}.  $G_F$ and $\alpha$ are the Fermi and the fine structure constant  at the $Z^0$ scale, respectively,   $V_{tb}$ and $V_{ts}$ are elements  of the Cabibbo-Kobayashi-Maskawa (CKM) matrix, and $\theta_W$ is  the Weinberg angle.
  The contribution of the operator with opposite chirality $O_R=({\bar b} s)_{V+A}({\bar \nu} \nu)_{V-A}$ is negligible.
The master function $X$ depends on the top quark mass $m_t$ and on the $W$ mass through the ratio $x_t=m_t^2/M_W^2$:
\be
X(x_t)=\eta_X \, X_0(x_t) \,\,\, .
\label{Xfunct}
\ee
The function $X_0$,
\be
X_0(x_t)=\frac{x_t}{8} \, \left[ \frac{x_t+2}{x_t-1}+\frac{3x_t - 6}{(x_t-1)^2} \log x_t \right] \,\,\, ,
\label{X0}
\ee
results  from the calculation of the loop (penguin and box) diagrams at leading order (LO) in $\alpha_s$ \cite{Inami:1980fz}, while the factor
 $\eta_X=0.994$  accounts for NLO $\alpha_s$ corrections \cite{Buchalla:1998ba}.
$X$ is flavour-universal and real, implying that, in  SM, it is possible to relate different modes with a neutrino pair in the final state, namely $B_d \to X_{s,d} \nu \bar \nu$ and $K^+ \to \pi^+ \nu \bar \nu$ or $K^0 \to \pi^0 \nu \bar \nu$.  Such  relations continue to hold   in NP models with minimal flavour violation.

The presence of a single operator in the Hamiltonian (\ref{heff}) makes the $b \to s \bar \nu \nu$  processes  easier to study in SM with respect to other rare decays described by a richer effective Hamiltonian, for instance  those induced by the $b \to s \ell^+ \ell^-$ transition. Moreover, long-distance effects threatening, e.g., the modes with charged leptons in the final state due to  hadron resonance contributions, are absent  in  modes into neutrino pairs.

 In general NP  extensions  the  new operator with opposite chirality $O_R$ can arise  and  the value of  $C_L^{SM}$ can be modified. The effective $b \to s \nu \bar \nu$ Hamiltonian is  given by
\be
H_{eff}=C_L O_L+ C_R O_R \,\,\, ,
\label{heffNP}
\ee
with   $C_{L,R}$  specific of the NP model. Notice that we only consider massless  left-handed neutrinos.

For the inclusive $B \to X_{s,d} \nu \bar \nu$ mode, the heavy quark mass expansion  allows to express the   decay rate  as a sum of terms  proportional to inverse  powers of the $b$ quark mass. The  ${\cal O}(\frac{1}{m_b^{2}})$ corrections are tiny, and the  same happens for the $q^2$ spectrum except for a small portion of the phase-space close to the kinematical end-point \cite{Falk:1995kn}.
For the  exclusive modes, in SM  a source of uncertainty is in the hadronic form factors describing the matrix element of the operator $O_L$
between the $B$ meson and  $K$ or $K^*$.  This problem  can  be circumvented in   $K \to \pi \nu \bar \nu$ modes, exploiting information on  the corresponding semileptonic modes (with one charged final lepton), and invoking isospin symmetry. On the other hand,
the  uncertainty  represented by the renormalization scale in  the QCD corrections is reduced by  the account  of NLO terms through the  $\eta_X$ factor  \cite{Buras:1998raa}.
Another difference with respect to  the analogous Kaon decay modes  is that in $B$ decays  the top quark contribution dominates, while in the Kaon  case,  namely the charged   $K^+ \to \pi^+ \nu \bar \nu$ decay, also the CKM enhanced  intermediate charm contribution has to be considered. This makes  the role of the $\alpha_s$ correction more important in the latter channel since  $\alpha_s(m_c)>\alpha_s(m_t)$.

In the study of NP effects  it is useful to introduce two parameters  \cite{Melikhov:1998ug},
\be
\epsilon^2=\frac{|C_L|^2+|C_R|^2}{|C_L^{SM}|^2} \,\,\, , \hskip 1 cm \eta=-\frac{{\rm Re}\left( C_L C_R^* \right)}{|C_L|^2+|C_R|^2} \,\,, \label{eta-eps}
\ee
which  probe deviations from  SM where $\left( \epsilon,\, \eta \right)_{SM}=(1,0)$.
In particular, $\eta$ is sensitive to the right-handed operator in the effective Hamiltonian, while $\epsilon$ mainly measures the deviation from  SM in the coefficient $C_L$.

\section{ $B \to K \nu \bar \nu$ and $B \to K^* \nu \bar \nu$}\label{Kmode}
The analysis of the exclusive $B \to K^{(*)} \nu \bar \nu$ modes requires  the hadronic matrix elements.
The $B \to K$  matrix element  can be parametrized in terms of two form factors,
\bea
&&<K(p^\prime)|{\bar s} \gamma_\mu b |B(p)>=
\nn \\&&=(p+p^\prime)_\mu
F_1(q^2) +{m_B^2-m_K^2 \over q^2} q_\mu \left
(F_0(q^2)-F_1(q^2)\right ) , \,\, \,\,\, \label{f0}
\eea
with $q=p-p^\prime$ and $F_1(0)=F_0(0)$.
Only $F_1$ is relevant for  decays to massless leptons.
Two dimensionless quantities can be defined,  the normalized neutrino pair invariant mass $s_B=q^2/m_B^2$,  and  the ratio ${\tilde m}_K=m_K/m_B$.
In  SM  the  decay distribution in  $s_B$ reads:
\be
\frac{d \Gamma^{SM}}{ds_B}= 3 \frac{|C_L^{SM}|^2}{96 \pi^3} m_B^5 \lambda^{3/2}(1,\,s_B,\,{\tilde m^2}_K)|F_1(s_B)|^2 \label{dgdsBKSM}\,\,,
\ee
with $C_L^{SM}$  in (\ref{heff}) and $\lambda(x,\,y,\,z)$
the triangular function.
In the  NP case this expression is generalized to
\be
\frac{d \Gamma}{ds_B}=3 \frac{|C_L+C_R|^2}{96 \pi^3} m_B^5 \lambda^{3/2}(1,\,s_B,\,{\tilde m^2}_K)|F_1(s_B)|^2 \label{dgdsBK}\,\,.
\ee
In both Eqs.~(\ref{dgdsBKSM}) and (\ref{dgdsBK})  the factor $3$  accounts for the sum over the three final neutrino flavours.
Modulo a factor of two, the distributions coincide with the distributions in $E_{miss}$,  the (missing) energy of the neutrino pair, since  $s_B=2x-1+{\tilde m}^2_K$, with   $x=E_{miss}/m_B$, and
\be
\frac{d \Gamma}{ds_B}=\frac{1}{2}\frac{d \Gamma}{dx} \label{dx}\,\,.
\ee

For the $B \to K^*$ matrix elements, we adopt the  usual parametrization in terms of form factors
\bea
<K^*(p^\prime,\epsilon)|{\bar s} \gamma_\mu (1-\gamma_5) b| B(p)>=\hspace*{3.5cm}  \nn \\
 \epsilon_{\mu \nu \alpha \beta} \epsilon^{* \nu} p^\alpha p^{\prime \beta}
{ 2 V(q^2) \over m_B + m_{K^*}}  \hspace*{0.5cm} \nn \\
- i \left [ \epsilon^*_\mu (m_B + m_{K^*}) A_1(q^2) -
(\epsilon^* \cdot q) (p+p')_\mu  {A_2(q^2) \over m_B + m_{K^*} }
\right. \,\,\,\,\nn\\
- \left. (\epsilon^* \cdot q) {2 m_{K^*} \over q^2}
\big(A_3(q^2) - A_0(q^2)\big) q_\mu \right ], \,\,\,\hspace*{0.5cm}   \label{a1}
\eea
where $\epsilon$ is the $K^*$ polarization vector.
The form factors are not all independent;  $A_3$ can be written as
\be
A_3(q^2) = {m_B + m_{K^*} \over 2 m_{K^*}}  A_1(q^2) - {m_B -
m_{K^*} \over 2 m_{K^*}}  A_2(q^2) ,
\ee
and  $A_3(0) = A_0(0)$. However,  $A_3$ and $A_0$ do not play a role in   transitions to massless leptons.

Three transversity amplitudes  can be defined, which depend either on $C_L-C_R$ or on $C_L+C_R$:
\begin{widetext}
\bea
{\cal A}_0(s_B)&=&-\frac{N(s_B)(C_L-C_R)}{{\tilde m}_{K^*} \sqrt{s_B}} \left[(1-{\tilde m}_{K^*}^2-s_B)(1+{\tilde m}_{K^*})A_1(s_B)-\lambda(1,{\tilde m^2}_{K^*},s_B)\frac{A_2(s_B)}{(1+{\tilde m}_{K^*})}\right] \nn \\
{\cal A}_\perp(s_B)&=&2 \sqrt{2} N(s_B) \lambda^{1/2}(1,{\tilde m^2}_{K^*},s_B)(C_L+C_R)\frac{V(s_B)}{(1+{\tilde m}_{K^*})} \label{transv-amp} \\
{\cal A}_\parallel(s_B)&=&-2 \sqrt{2} N(s_B) (C_L-C_R)(1+{\tilde m}_{K^*})A_1(s_B) \,\,\,\, , \nn
\eea
\end{widetext}
with  ${\tilde m}_{K^*}=m_{K^*}/m_B$ and the  function $N(s_B)$  defined as  $N(s_B)={\left[\displaystyle\frac{m_B^3 s_B \lambda^{1/2}(1,{\tilde m^2}_{K^*},s_B)}{3\cdot 2^7 \,\pi^3}\right]}^{1/2}$.
The differential distributions in $s_B$ for a longitudinally or transversely polarized $K^*$ (with helicity $h=+1$ or $h=-1$) can be written in terms of these amplitudes.
Exploiting the definitions (\ref{eta-eps}),  one finds for the
sum over the three neutrino flavours:
\begin{widetext}
\bea
\frac{d \Gamma_L}{ds_B}&=&3 m_B^2 {\cal A}_0^2= \left(\frac{d \Gamma_L}{ds_B}\right)_{SM} \epsilon^2 \,(1+2 \eta)\nn \\
\frac{d \Gamma_\pm}{ds_B}&=&\frac{3}{2} m_B^2|{\cal A}_\perp \mp {\cal A}_\parallel |^2 \nn \\
\frac{d \Gamma_T}{ds_B}&=&\frac{d \Gamma_+}{ds_B}+ \frac{d \Gamma_-}{ds_B}=
3 m_B^2 \left({\cal A}_\perp^2+{\cal A}_\parallel^2\right) =\left(\frac{d \Gamma_T}{ds_B}\right)_{SM} \epsilon^2 \,\left(1+2 \eta \, f_T(s_B)\right)
\label{spectra} \\
\frac{d \Gamma}{ds_B}&=&3 m_B^2 \left({\cal A}_0^2+{\cal A}_\perp^2+{\cal A}_\parallel^2 \right) =\left(\frac{d \Gamma}{ds_B}\right)_{SM} \epsilon^2
\,\left(1+2 \eta \, f(s_B)\right) \,\,\,\, , \nn
\eea
with
\bea
f_T(s_B)&=&\frac{(1+{\tilde m}_{K^*})^4[A_1(s_B)]^2-\lambda [V(s_B)]^2}{(1+{\tilde m}_{K^*})^4[A_1(s_B)]^2+\lambda [V(s_B)]^2} \nn \\
f(s_B)&=& \frac{\left[ (1+{\tilde m}_{K^*})^2(1-s_B-{\tilde m}_{K^*}^2)A_1(s_B)-\lambda A_2(s_B) \right]^2+8 {\tilde m}_{K^*}^2 s_B \left[ (1+{\tilde m}_{K^*})^4[A_1(s_B)]^2-\lambda [V(s_B)]^2 \right]}{\left[ (1+{\tilde m}_{K^*})^2(1-s_B-{\tilde m}_{K^*}^2)A_1(s_B)-\lambda A_2(s_B) \right]^2+8 {\tilde m}_{K^*}^2 s_B \left[ (1+{\tilde m}_{K^*})^4[A_1(s_B)]^2+\lambda[V(s_B)]^2 \right]} \,\,\, . \,\,\,\,\,\,\,\, \label{Rterms}
\eea
\end{widetext}
In Eq.~(\ref{Rterms}) we use the notation  $\lambda=\lambda(1,{\tilde m^2}_{K^*},s_B)$; the factor 3 in Eqs.~(\ref{spectra}) accounts for the sum over the neutrino species.
Also in this case,  the distributions in $s_B$ can be converted in neutrino missing energy distributions using Eq.~(\ref{dx}).

Starting from  the above defined quantities, several observables can be constructed.

 The polarization fractions $F_{L,T}$ can  be considered  \cite{Altmannshofer:2009ma},
\be
\frac{d F_{L,T} }{ds_B}=\frac{d \Gamma_{L,T}/ds_B}{d \Gamma/ds_B} \label{dFLT}
\ee
in which several  hadronic and parametric uncertainties are  reduced or  even canceled (namely the overall quantities, like the CKM elements in SM).
The integrated polarization fractions  can be obtained,  integrating separately    the numerator and the denominator in Eq.~(\ref{dFLT}):
\be
F_{L,T}=\frac{1}{\Gamma} \, \int_0^{1-{\tilde m}_{K^*}^2} ds_B \, \frac{d F_{L,T} }{ds_B} \,\,.
\label{FLTint}
\ee

Another observable is  the ratio of branching fractions involving  $K$ and the transversely polarized $K^*$  \cite{Melikhov:1998ug},
\bea
R_{K/K^*}=\frac{{\cal B}(B \to K \,\nu \bar \nu)  }{{\cal B}(B \to K^*_{h=-1} \,\nu \bar \nu) +{\cal B}(B \to K^*_{h=+1} \,\nu \bar \nu) } \,\, ,\nonumber  \\
\label{RK-KstarT}
\eea
which is  sensitive to  $\eta$.

In \cite{Melikhov:1998ug} the transverse asymmetry has been proposed
\be
A_T=\frac{{\cal B}(B \to K^*_{h=-1} \, \nu \bar \nu) -{\cal B}(B \to K^*_{h=+1} \, \nu \bar \nu) }{{\cal B}(B \to K^*_{h=-1} \,\nu \bar \nu) +{\cal B}(B \to K^*_{h=+1} \,\nu \bar \nu) } \,\, ,
\label{asymT}
\ee
for which   a reduced  hadronic uncertainty is expected. However, its measurement would require the determination of the lepton pair polarization \cite{Egede:2008uy},
therefore we consider it  only for a theoretical analysis.

The observables can probe NP effects, as the ones envisaged in warped five-dimensional extensions of the standard model.

\section{Randall-Sundrum model with custodial protection}\label{sect:RS}
The  motivation of the Randall-Sundrum (RS) model is the possibility of addressing, among others,  the hierarchy and the flavour problems invoking the same geometrical mechanism\cite{RandallSundrum}. For a  description of the model,  in particular for the flavour phenomenology, we refer to  \cite{Albrecht:2009xr}. Here we
 briefly illustrate the main features  of the custodially-protected RS$_c$ model, adopting  the same notations  of  our  analysis of  $B \to K^* \ell^+ \ell^-$  in this framework  \cite{Biancofiore:2014wpa},   with the parameter space determined there.

The RS$_c$ model is a new  physics scenario in which the spacetime is supposed to be five-dimensional with coordinates $(x,\,y)$, $x$ being the ordinary 4D Minkowskian coordinates, and metric
\bea
ds^2&=&e^{-2 k y} \eta_{\mu \nu} dx^\mu dx^\nu - dy^2 \,\,\, , \nn \\
\eta_{\mu \nu}&=&diag(+1,-1,-1,-1) \,\,\, . \label{metric}
\eea
 The (fifth) coordinate $y$ varies in the range $0 \le y \le L$;  $y=0$ is  identified with  the so-called UV brane,   $y=L$ with the IR brane.
To address the hierarchy problem,  the parameter $k$ in the metric (\ref{metric}) is chosen  $k\simeq {\cal O}(M_{Planck})$: specifically, $k$ is set to  $k=10^{19}$ GeV.
We adopt the variant of  the model  based on the  gauge group
\be
SU(3)_c \times SU(2)_L \times SU(2)_R \times U(1)_X \times P_{L,R} \,\,
\label{group}
\ee
which, together with the metric,    defines the Randall-Sundrum model with custodial protection RS$_c$ \cite{contino,carena,cacciapaglia}.
Indeed, the action of the discrete $Z_2$ $P_{L,R}$ symmetry,  implying a mirror action of the two $SU(2)_{L,R}$ groups,  guarantees the custodial protection avoiding
 large $Z$ couplings to left-handed fermions, experimentally not allowed.

Appropriate boundary conditions (BC) on the UV brane permit to break the gauge group (\ref{group}) to the SM gauge group, which further undergoes a
 spontaneous symmetry breaking through a Higgs mechanism,   as in SM.
Among the various SM fields, the Higgs one is chosen to be localized close to the IR brane, while all the other fields
can propagate in the bulk. Here we consider a Higgs field completely localized at $y=L$.

The  existence of a compact fifth dimension leads to a tower of Kaluza-Klein (KK) excitations for all particles. As customary in extra-dimensional models,  particles having a SM correspondent  can be distinguished from those without SM partners by  the choice of their field boundary conditions, so  that only for some choices  a zero mode in the KK mode expansion exists.
Two choices for  BC are adopted: Neumann BC  on both branes (++), or Dirichlet BC on the UV brane and Neumann BC on the IR one (-+). The zero modes exist only for  fields with  (++) BC, and are identified with  the SM particles.
The KK decomposition  has the  general form
\be
F(x,y)=\frac{1}{\sqrt{L}}\sum_k F^{(k)}(x) f^{(k)}(y)\,\,.
\label{genericKK}
\ee
For each field $F(x,y)$  the functions $f^{(k)}(y)$  are  referred to  as the 5D field profiles, and $F^{(k)}(x)$ are the effective 4D fields.
The 5D profiles are obtained from  the 5D Lagrangian densities for the various fields,   solving the resulting 5D equations of motion. This  can be performed before the EWSB takes place \cite{Albrecht:2009xr}.
Afterwards,  the ratio $v/M_{KK}$ of the Higgs vacuum expectation value (vev) $v$ and the  mass  of the lowest KK mode $M_{KK}$ is treated as a perturbation. The effective 4D Lagrangian is derived integrating over $y$, and the Feynman rules follow after  the  neglect of  terms of ${\cal O}(v^2/M_{KK}^2)$, or higher. The mixing  occuring  between  SM fermions and  higher KK fermion modes is neglected, being ${\cal O}(v^2/M_{KK}^2)$.
In the case of gauge bosons,   modes up to the first KK excitation (1-mode) are taken into account \cite{Albrecht:2009xr}.

Among the  particles without  a SM counterpart, new gauge bosons are predicted to exist, due to the enlarged  gauge group.
The gauge bosons of $SU(2)_L$ and $SU(2)_R$ are denoted by $W_L^{a,\mu}$  and $W_R^{a, \mu}$  ($a=1,2,3$),  respectively;   the gauge choices $W_{L,R}^{a,5}=0$ and $\partial_\mu W_{L,R}^{a,\mu}=0$   are adopted, as  for all the other gauge bosons. The equality $g_L=g_R=g$ for the  $SU(2)_{L,R}$ gauge couplings  is a consequence of the  $P_{L,R}$ symmetry .

The  eight gauge fields corresponding to $SU(3)_c$ remain identified with the gluons as in SM, while  a new gauge field  $X_\mu$, from   the $U(1)_X$, has  coupling  $g_X$.
All the 5D couplings are dimensionful, and are connected  to their 4D counterparts  by the relation
$g^{4D}={g^{5D}}/{\sqrt{L}}$.

A mixing occurs among the various gauge fields. Charged gauge bosons are defined as in  SM:
\be
W_{L(R)\mu}^\pm=\frac{W^1_{L(R)\mu} \mp i W^2_{L(R)\mu}}{\sqrt{2}}\,\,.
\label{chargedW}
\ee
On the other hand, $W_R^3$ and $X$  mix through an angle $\phi$. The resulting fields are  $Z_{X }$ and $B$; the latter mixes with
$W_{L \,}^3$  with an angle $\psi$, providing the $Z$ and $A$ fields as in  SM.

In summary,  the gauge boson content of the model,    together with the BC, is:  eight gluons $G_\mu$ with BC $(++)$,  four charged bosons $W_L^\pm (++)$ and $W_R^\pm (-+)$,
three  neutral bosons $A(++)$, $Z(++)$ and $Z_{X}(-+)$.
For each of these vector fields, the KK expansion  is
\be
V_\mu(x,y)=\frac{1}{\sqrt{L}}\sum_{n=0}^{\infty}V_\mu^{(n)}(x) f_V^{(n)}(y) \,\,. \label{KK-vec-bos}
\ee
The profiles of the zero-modes are flat, $f_V^{(0)}(y)=1$. As
 for the 1-modes,  for gauge bosons having a zero-mode they are denoted by $g(y)$ and their mass is denoted as $M_{++}$;
 for gauge bosons without a zero-mode, they are indicated by ${\tilde g}(y)$, with mass   $M_{-+}$.
We refer  to the Appendix of \cite{Biancofiore:2014wpa} for the expressions of these quantities and for the notation.
The  solution of the equation of motion provides $M_{++} \simeq 2.45 f$ and $M_{-+} \simeq 2.40 f$, where  $f$ is the  dimensionful parameter   $f=k \, e^{-kL}$.  We set this parameter   to $f=1$ TeV,  coherently with  other studies \cite{buras1,buras4,Blanke:2012tv}.

Before the EWSB the zero modes of the gauge bosons (if present) are massless, while higher KK excitations are massive.
 Since the two groups $SU(3)$ (for QCD) and $U(1)_{em}$ remain unbroken, the zero modes of  gluons and  photon are massless as in  SM, but
 their  KK excitations are massive.

Mixing  also occurs  among zero modes and higher KK modes of gauge fields. Neglecting modes with KK number larger than $1$, the mixing involves
 the charged bosons $W_L^{\pm(0)},\,W_L^{\pm(1)}$ and $W_R^{\pm(1)}$, with the result
\be
\left(\begin{array}{c} W^\pm \\W_H^\pm \\ W^{\prime \pm} \end{array} \right)={\cal G}_W \,\,\left(\begin{array}{c} W_L^{\pm (0)} \\W_L^{\pm (1)} \\ W_R^{ \pm (1)} \end{array} \right) \,\,\, ,
\ee
and the neutral bosons $Z^{(0)}$, $Z^{(1)}$ and $Z_X^{(1)}$  according to the  pattern:
\be
\left(\begin{array}{c} Z \\Z_H \\ Z^{\prime} \end{array} \right)={\cal G}_Z \,\,\left(\begin{array}{c} Z^{ (0)} \\Z^{ (1)} \\ Z_X^{  (1)} \end{array} \right) \,\,.
\ee
 The  expressions of the matrices ${\cal G}_W$ and ${\cal G}_Z$ and the  masses of the mass eigenstates can be found in Ref.~\cite{Albrecht:2009xr}.

Moving to the Higgs sector, the Higgs field
$H(x,y)$  transforms as a bidoublet under $SU(2)_L \times SU(2)_R$ and as a singlet under $U(1)_X$. It contains two charged and  two neutral components:
\be
H(x,y)=\left(\begin{array}{cc}
\frac{\pi^+}{\sqrt{2}} & -\frac{h^0-i\pi^0}{2} \label{Hbidoublet} \\
\frac{h^0+i\pi^0}{2}  & \frac{\pi^-}{\sqrt{2}}
\end{array}\right)  \,\,\, . \nn
\ee
Its KK decomposition  reads
\be
H(x,y)=\frac{1}{\sqrt{L}}\sum_k H^{(k)}(x) h^{(k)}(y)\,\,.
\label{higgsKK}
\ee
The  localization  on the IR brane leads to the profile
\be
h(y)\equiv h^{(0)}(y)  \simeq e^{kL} \delta(y-L) \,\,.
\label{H-loc}
\ee
Furthermore, one chooses that
 only the neutral field  $h^0$ has  a non-vanishing  vacuum expectation value   $v=246.22$ GeV, as in  SM.

The most involved sector is the fermion one.
We refer  to \cite{Albrecht:2009xr} for the description of  the fermion representations.
Here, we  mention that, considering three generations of quarks and leptons, SM left-handed doublets are collected  in a bidoublet of $SU(2)_L \times SU(2)_R$, together with two new fermions.
 Right-handed up-type quarks are singlets, while no corresponding fields exist in the case of leptons, for  left-handed neutrinos.
  Right-handed down-type quarks, as well as  charged leptons are  in multiplets  transforming as $(3,1) \oplus (1,3)$ under  $SU(2)_L \times SU(2)_R$, and   additional new fermions are also present in such multiplets.
   The electric charge is related to the third component of the $SU(2)_L$ and $SU(2)_R$ isospins and to the charge $Q_X$ through the equation $Q=T^3_L+T^3_R+Q_X$.

The presence of new fermions will not affect our analysis, since
  we  only take into account  the zero-modes of SM quarks and leptons.
The zero-mode profiles are obtained
solving the equations of motion for ordinary fermions, with  result  denoted as $f_{L,R}^{(0)}(y,c)$:
\be
f^{(0)}(y,c)=\sqrt{\frac{(1-2c)kL}{e^{(1-2c)kL}-1}}e^{-cky} \,\,.
\ee
The difference between right- and left-handed fermion profiles lies in the parameter $c$,    the fermion mass in the bulk.
Fields belonging to the same $SU(2)_L \times SU(2)_R$ multiplet share the same value of $c$, as  is the case for $u_L$ and $d_L$, $c_L$ and $s_L$, $t_L$ and $b_L$,  as well as for $\nu_\ell$ and $\ell^-_L$ ($\ell=e,\mu,\tau$). We choose real  $c$ parameters.

Other parameters of the model enter  when considering the quark mixing. As  in  SM, quark mass eigenstates are obtained by a rotation of flavour eigenstates. The rotation matrices of up-type left (right) and down-type left (right) quarks are denoted by ${\cal U}_{L(R)}$, ${\cal D}_{L(R)}$, respectively. Moreover, the CKM matrix is obtained as  $V_{CKM}={\cal U}_L^\dagger {\cal D}_L$.
At odds with  SM, in which the presence of the CKM matrix affects only charged current interactions, in RS$_c$  the rotation matrices  also affect neutral current interactions,  and this leads to the occurrence of flavour changing neutral currents at tree level mediated by the three neutral EW gauge bosons $Z,\,Z^\prime, \, Z_H$,  as well as by the first KK mode of the photon and of the gluon (however, gluons play no role in processes with leptons in the final state, and photons do not contribute to the transitions to neutrinos).  The corresponding Feynman rules involve the overlap integrals of fermion and gauge boson profiles,
\bea
{\cal R}_{f_i f_j} &=& \frac{1}{L} \int_0^L dy  \, e^{ky} \,  f_{f_i}^{(0)}(y,c_i) \, f_{f_j}^{(0)}(y,c_j) \, g(y) \nn \\
\tilde{\cal R}_{f_i f_j} &=& \frac{1}{L} \int_0^L dy  \,e^{ky} \,  f_{f_i}^{(0)}(y,c_i) \, f_{f_j}^{(0)}(y,c_j) \, {\tilde g}(y) \,\, ,
 \eea
collected in two  matrices
${\cal R}_f=diag \left({\cal R}_{f_1 f_1}, {\cal R}_{f_2 f_2}, {\cal R}_{f_3 f_3}\right)$ and ${ \tilde{\cal R}}_f=diag \left(\tilde{\cal R}_{f_1 f_1}, \tilde{\cal R}_{f_2 f_2}, \tilde{\cal R}_{f_3 f_3}\right)$.
After the rotation to mass eigenstates, the quantities  appearing in the Feynman rules of the model are the products  ${\cal M}^\dagger {\cal R}_f {\cal M}$, where ${\cal M}={\cal U}_{L,R}, {\cal D}_{L,R}$.
The details, as well as the list of Feynman rules, can be found in \cite{Albrecht:2009xr,Biancofiore:2014wpa}.

The required elements of the rotation matrices   can be written in terms
of the quark profiles, and of the 5D Yukawa couplings   denoted by $\lambda_{ij}^u$ for up-type quarks and $\lambda_{ij}^d$ for down type quarks, respectively.
The effective 4D Yukawa couplings are given by
\be
Y_{ij}^{u(d)}=\frac{1}{\sqrt{2}}\frac{1}{L^{3/2}} \int_0^L \, dy \, \lambda_{ij}^{u(d)} f_{q_L^i}^{(0)}(y)f_{u_R^j(d_R^j)}^{(0)}(y) h(y)
\label{Yud}
\,\,.
\ee
This relation produces  the  fermion mass and mixing hierarchy, due to the exponential dependence of  the fermion profiles  on the bulk mass parameters  \cite{gherghetta,grossman-neubert}.

The elements of the  matrices ${\cal U}_{L(R)}$ and ${\cal D}_{L(R)}$ are not all independent, not only because the constraint $V_{CKM}={\cal U}_L^\dagger {\cal D}_L$ must be fulfilled, but also because  the Yukawa couplings determine the quark masses.
In particular, the following relations must be satisfied:
\bea
m_u&=& \frac{v}{\sqrt{2}}\frac{det(\lambda^u)}{\lambda^u_{33}\lambda^u_{22}-\lambda^u_{23}\lambda^u_{32}}\frac{e^{kL}}{L}f_{u_L}f_{u_R} \nn \\
m_c &=& \frac{v}{\sqrt{2}} \frac{\lambda^u_{33}\lambda^u_{22}-\lambda^u_{23}\lambda^u_{32}}{\lambda^u_{33}}\frac{e^{kL}}{L}f_{c_L}f_{c_R} \label{u-masses} \\
m_t&=&\frac{v}{\sqrt{2}}\lambda^u_{33}\frac{e^{kL}}{L}f_{t_L}f_{t_R}\,\,\, , \nn \eea
as well as the analogous relations for down-type quarks with the replacement $\lambda^u \to \lambda^d$
(with the  notation  $f_{q_{L,R}}=f_{q_{L,R}}^{(0)}(y=L,c_{q_{L,R}})$).

In our  analysis we adopt simplifying assumptions, such as considering  real  entries of the matrices $\lambda^{u,d}$. As a consequence,
  after  the quark mass constraints have been imposed, there are six  independent entries among the elements of the Yukawa matrices, which  we choose \footnote{A  parametrization of the matrices $\lambda^{u,d}$ with complex entries is described in \cite{buras1}.}
\bea
\lambda^u_{12} \,\,\, , \hskip 0.5 cm  \lambda^u_{13} \,\,\, ,\hskip 0.5 cm \lambda^u_{23} \,\,\, ,\nn \\
\lambda^d_{12} \,\,\, ,\hskip 0.5 cm  \lambda^d_{13} \,\,\, ,\hskip 0.5 cm \lambda^d_{23}  \,\,\, .
\label{lambdas}
\eea
Therefore, the set of  input parameters in our analysis is composed by the six quantities in (\ref{lambdas}), together with  the bulk mass parameters.
Before describing our strategy for the numerical study, we  discuss  the Wilson coefficients in the effective Hamiltonian (\ref{heffNP}) in RS$_c$, and how they are modified  with respect to the standard model.

\section{Effective  $b \to s \nu \bar \nu$ Hamiltonian in RS$_c$ model}\label{sect:WRS}
In SM the  Wilson coefficients of the left- and right-handed operators $O_L$ and $O_R$ in  the effective Hamiltonian (\ref{heff}),(\ref{heffNP}) are given by
\bea
C_L^{SM}&=&{G_F \over \sqrt{2}} {\alpha \over 2 \pi \sin^2\theta_W} V_{tb}^* V_{ts} X(x_t) \label{CLSM} \\
C_R^{SM}&=&0 \nn \,\,.
\eea
 These coefficients are modified in  the RS$_c$, in which a right-handed operator   $O_R$ is  present:
\bea
C_L^{RS}&=&{G_F \over \sqrt{2}} {\alpha \over 2 \pi \sin^2\theta_W} V_{tb}^* V_{ts} X_L^{RS} \label{CLRS} \\
C_R^{RS}&=&{G_F \over \sqrt{2}} {\alpha \over 2 \pi \sin^2\theta_W} V_{tb}^* V_{ts} X_R^{RS}  \label{CRRS} \,\,,
\eea
with $X_L^{RS}=X(x_t)+\Delta X_L$ and
\bea
\Delta X_L &=& \frac{1}{V_{tb}V_{ts}^*}\sum_{X=Z,\,Z^\prime,\,Z_H}\frac{\Delta_L^{bs}(X) \Delta^{\nu \bar \nu} (X)}{4M_X^2 g_{SM}^2}
\label{DeltaXL} \\
X_R^{RS}&=&\frac{1}{V_{tb}V_{ts}^*}\sum_{X=Z,\,Z^\prime,\,Z_H}\frac{\Delta_R^{bs}(X) \Delta^{\nu \bar \nu} (X)}{4M_X^2 g_{SM}^2} \,\,\, . \label{XR}
\eea
The constant $g_{SM}^2$ is defined as    $\dd g_{SM}^2=\frac{G_F}{\sqrt{2}}\frac{\alpha}{2 \pi \sin^2(\theta_W)}$.
 $\Delta_{L,R}^{f_i f_j}(X)$ is the coupling of a  gauge boson $X$ to a pair of fermions $f_i f_j$;  it  can be read from the Feynman rules described in the Appendix of Ref.~\cite{Biancofiore:2014wpa}.

The new  contributions can be evaluated  scanning  the  parameter space of the RS$_c$ model.
As  described in  \cite{Biancofiore:2014wpa}, we  require that
the elements of the matrices  $\lambda^{u,d}$  lie in a range  assuring the perturbativity of the model up to the scale of the first three KK modes: $|\lambda^{d,u}_{ij}| \le 3/k$. Moreover, the diagonal elements of such matrices are fixed imposing the quark mass constraints.
The six remaining parameters, those in (\ref{lambdas}),   must be fixed together with the bulk mass parameters for the quarks,
enforcing  quark mass and CKM constraints.
Imposing for  the quark masses the values obtained at the scale ${\cal O}(M_{KK})$ through renormalization group evolution, starting from
\be
m_d= 4.9 \,\, {\rm MeV} , \hskip 0.1 cm m_s=90 \,\, {\rm MeV}  ,  \hskip 0.1  cm  m_b= 4.8 \,\, {\rm GeV}  \,\,\, ,
\ee
the  following quark mass bulk parameters have been  determined  \cite{Duling:2010lqa}:
\bea
c_L^{u,d}=0.63 \,\, , \hskip 0.3 cm   c_L^{c,s}=0.57 \,\, ,\hskip 0.3 cm c_L^{b,t}\in [0.40,\, 0.45] \,\, , \,\,\, \nn \\
c_R^u = 0.67 \,\, ,\hskip 0.3 cm c_R^c=0.53 \,\, ,\hskip 0.3 cm c_R^t=-0.35 \,\, , \,\,\,\,\, \label{bulkmasses}  \\
c_R^d = 0.66 \,\, , \hskip 0.3 cm c_R^s=0.60 \,\, ,\hskip 0.3 cm c_R^b=0.57 \,\,\,\,. \,\,\,\,\,\,\,\,\, \nn
\eea
We quote a range in the case of  the left-handed doublet of the third quark generation,  since further  constraints are imposed, i.e. those derived in \cite{neubertRS} using  the  measurements of  the coupling $Z{\bar b}b$, of the $b$-quark left-right asymmetry parameter and of the forward-backward asymmetry for $b$ quarks  \cite{ALEPH:2005ab}.

For leptons,  the bulk masses are set to $c_\ell=0.7$   \cite{buras2}. Other numerical determinations  of the fermion  bulk mass parameters can be found in \cite{Huber:2000ie,burdman,agashe,Agashe:2004ay,Fitzpatrick:2007sa,neubertRS,Archer:2011bk}.

In correspondence to the values fixed above, we generate the six $\lambda$ parameters in Eq.~(\ref{lambdas}) imposing the CKM constraints.
Specifically, we require $|V_{cb}|$ and $|V_{ub}|$ in the largest range found from their experimental determinations from inclusive and exclusive $B$ decays \cite{Amhis:2012bh}, and
 that $|V_{us}|$ lies within $2\%$ of the central value reported by the Particle Data Group \cite{Beringer:1900zz}. Hence, the selected ranges are:
\bea
|V_{cb}| &\in& [0.038 - 0.043]  \nn \\
 |V_{ub} |&\in& [0.00294 - 0.00434]    \\
| V_{us} |&\in& [0.22 - 0.23] \,\,\, . \nn
\label{VxbVus}
\eea
The parameter space is further restricted, as in \cite{Biancofiore:2014wpa},   imposing that  the
$B \to K^* \mu^+ \mu^-$ and $B \to X_s \gamma$ branching fractions lie within the
$2\sigma$ range of  the measurements
\bea
\hskip -0.2cm {\cal B}(B \to K^* \mu^+ \mu^-)_{exp} &=& (1.02 \pm^{0.14}_{0.13} \pm0.05)\times 10^{-6} \, ,\,\,\,\,\,  \label{Kstarmumuexp} \\
\hskip -0.2cm{\cal B}(B \to X_s \gamma)_{exp} &=& (3.43 \pm 0.21 \pm 0.07) \times 10^{-4} . \,\,\,\,\,\,\, \label{BXsgammaexp}
\eea
The datum   (\ref{Kstarmumuexp}) is the BaBar average of the branching fractions of the four modes $B^{+,0} \to K^{* +,0} \mu^+ \mu^- $ and  $B^{+,0} \to K^{* +,0} e^+ e^-$  \cite{Lees:2012tva},
while the value in  (\ref{BXsgammaexp}) is  the HFAG Collaboration average for this inclusive rare radiative $B$ decay width \cite{Amhis:2012bh}.
With the selected set of points in the parameter space it is also possible to reproduce in  the RS$_c$ model, using the expressions in   \cite{buras1},  the mass difference of the neutral $B_s$ mesons $\Delta M_s$ within $20\%$ of  the central value of the experimental measurement  $\Delta M_s=17.69$ ps$^{-1}$ \cite{Amhis:2012bh} .

\begin{figure}[t!]
\includegraphics[width = 0.4\textwidth]{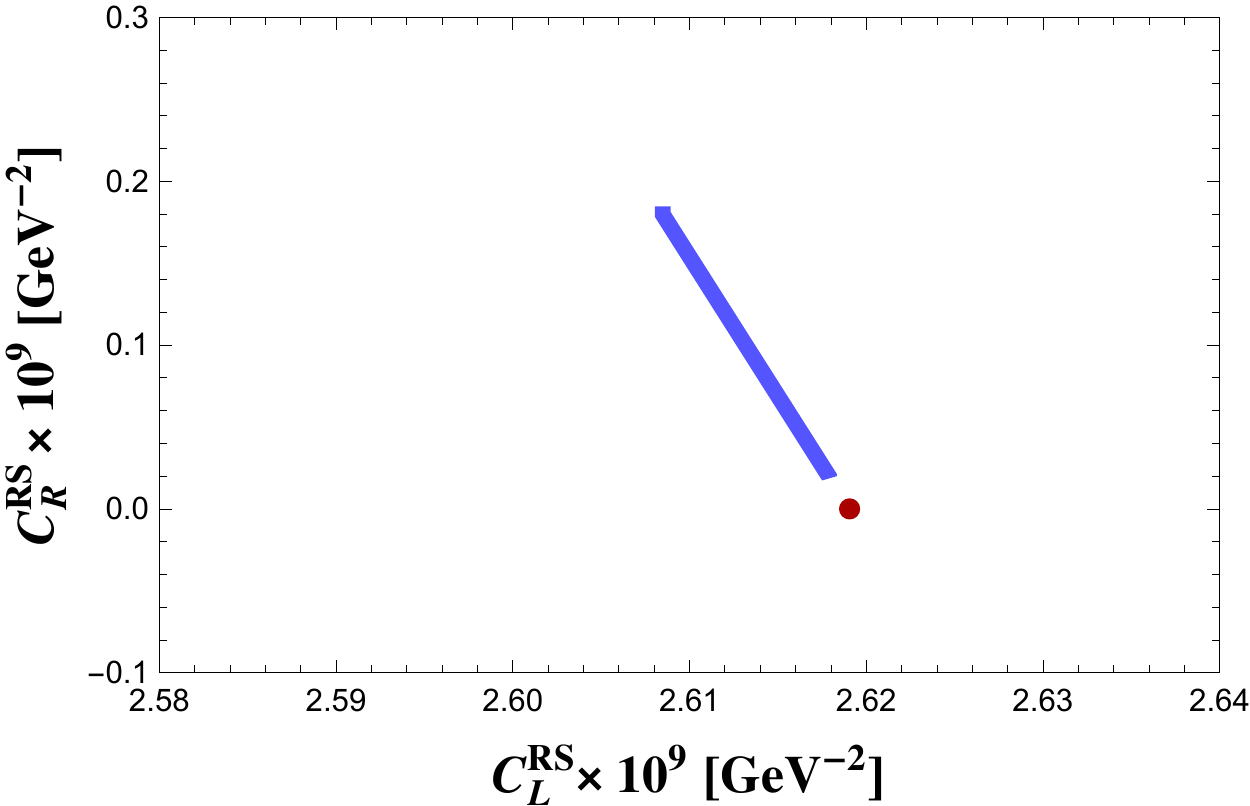}
\caption{Correlation between $C_R^{RS}$ and  $C_L^{RS}$ in the RS$_c$ model  (blue curve). The red dot corresponds to the central  SM values.}\label{fig:CLR}
\end{figure}

\begin{figure}[t!]
\includegraphics[width = 0.4\textwidth]{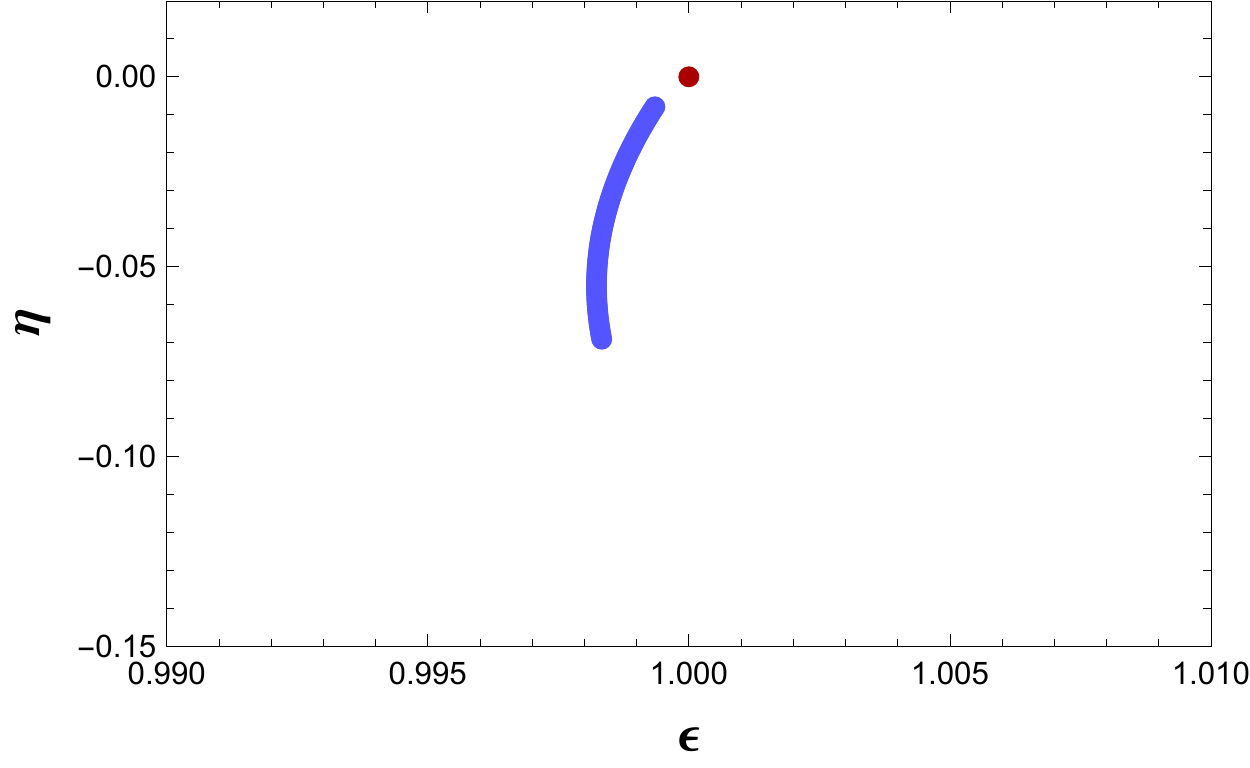}
\caption{Correlation between the parameters $\eta$ and  $\epsilon$, defined in (\ref{eta-eps}),  in the RS$_c$ model (blue curve). The red dot corresponds to SM.}\label{fig:etaeps}
\end{figure}

Scanning the parameter space resulting from  all the   constraints, we obtain the coefficients  $C_{L}^{RS}$ and  $C_{R}^{RS}$ and their correlation, as shown
in Fig.~\ref{fig:CLR}. The  resulting parameters $\eta$ and $\epsilon$, defined in (\ref{eta-eps}),  are depicted in Fig.~\ref{fig:etaeps}.
The first observation concerns  the  right-handed coupling:  we find that
a  deviation from SM is predicted,  with the  maximum value   $C_R^{RS}= 0.186\times10^{-9}$ GeV$^{-2}$. For the  left-handed coupling we obtain the maximum $\Delta C_L=C_L^{RS}-C_L^{SM}=-0.011\times10^{-9}$ GeV$^{-2}$. The largest deviation of  $\eta$ from its SM value $\eta=0$ is $\eta=-0.075$. As a signature of the model, $C_L$ and $C_R$ are  anticorrelated,
as shown Fig.~\ref{fig:CLR},  and this has a definite impact on the various observables that we are going to discuss in details.

\section{$B \to K \nu \bar \nu$  and $B \to K^{*} \nu \bar \nu$  observables  in RS$_c$}\label{sect:BKstar}
To compare the  RS$_c$ predictions  to the SM results for the exclusive $B \to K \nu \bar \nu$  and $B \to K^{*} \nu \bar \nu$ decay observables defined in section \ref{Kmode}  we need  the $B \to K^{(*)}$ form factors. Here   we use the  light-cone QCD sum rule determination  \cite{Ball:2004rg}. Other QCD sum rule determinations, as the one in \cite{Colangelo:1995jv} from three-point correlation functions,  have  larger uncertainties.  The one in \cite{Altmannshofer:2008dz},  which includes  QCD factorization corrections, only provides  the $B \to K^*$ matrix elements, while  we need the full set of $B \to K$,  $B \to K^*$, as well as  $B_s \to  \phi$ matrix elements.  Lattice QCD results are now available \cite{Bouchard:2013eph,Horgan:2013hoa}, and we  comment below on the  differences.
\begin{figure}[b!]
\includegraphics[width = 0.4\textwidth]{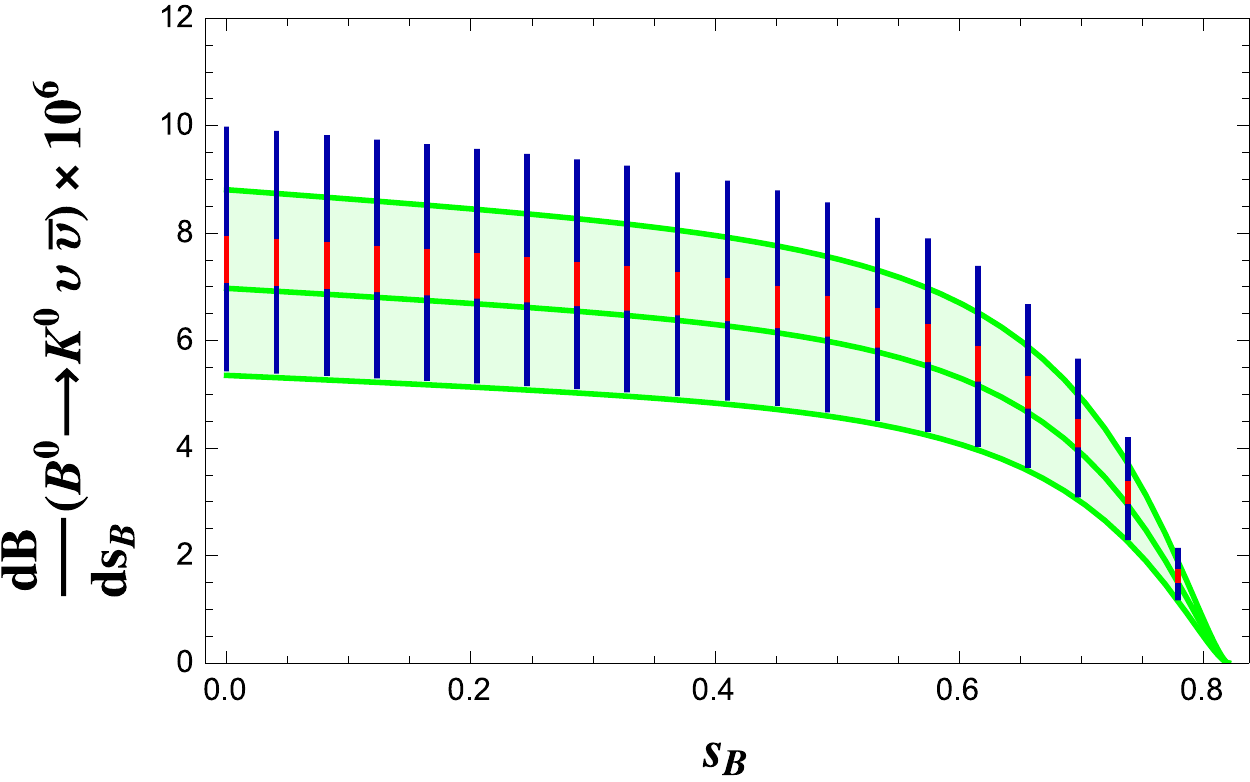}
\caption{$\displaystyle \frac{d {\cal B}}{ds_B}(B^0 \to K^0 \nu \bar \nu)$ distribution  in SM, including the uncertainty on the form factor $F_1(0)$  (green region), and  in RS$_c$  for the central value of $F_1(0)$  (red points) and including the uncertainty of the form factor at $s_B=0$ (blue bars). }\label{fig:dGKdsB}
\end{figure}

In Fig.~\ref{fig:dGKdsB} we depict the differential distribution $\frac{d {\cal B}}{ds_B}(B^0 \to K^0 \nu \bar \nu)$   in the whole kinematical range  $0\le s_B \le  \left(1-\frac{m_K}{m_B} \right)^2 $ in  SM, including the uncertainty on the form factor $F_1(0)$ quoted in \cite{Ball:2004rg} and using  the measured lifetime    $\tau(B^0)=1.519\pm 0.005$~ps~\cite{Amhis:2012bh}.
The   predicted branching fraction
\be
{\cal B}(B^0 \to K^0 \nu \bar \nu)_{SM}=(4.6 \pm1.1)\, \times 10^{-6} \,\,\label{BRKSMBall}
\ee
has a larger uncertainty than the one in (\ref{LCSR}), due to our more conservative errors on the  form factors.
The modifications in RS$_c$,  obtained for the central value of $F_1(0)$ and  accounting for  the uncertainty on $F_1(0)$,  are also shown in Fig.~\ref{fig:dGKdsB}  and produce a prediction for the  branching fraction  spanning a somewhat wider range,
\be
{\cal B}(B^0 \to K^0 \nu \bar \nu)_{RS}\in[3.45 - 6.65]\, \times 10^{-6} \,\,.\label{BRKRSBall}
\ee
A similar result is obtained for the charged mode.  Hence, the  present experimental upper bounds require an  improvement  by a factor of 3-4  in the case of BaBar, Eq.~(\ref{Babar}), and of about one order of magnitude in the case of Belle,   Eq.~(\ref{belle}),  to become sensitive to these processes, a  task within the possibilities of   high-luminosity facilities  such as Belle II.

\begin{figure}[b]
\includegraphics[width = 0.4\textwidth]{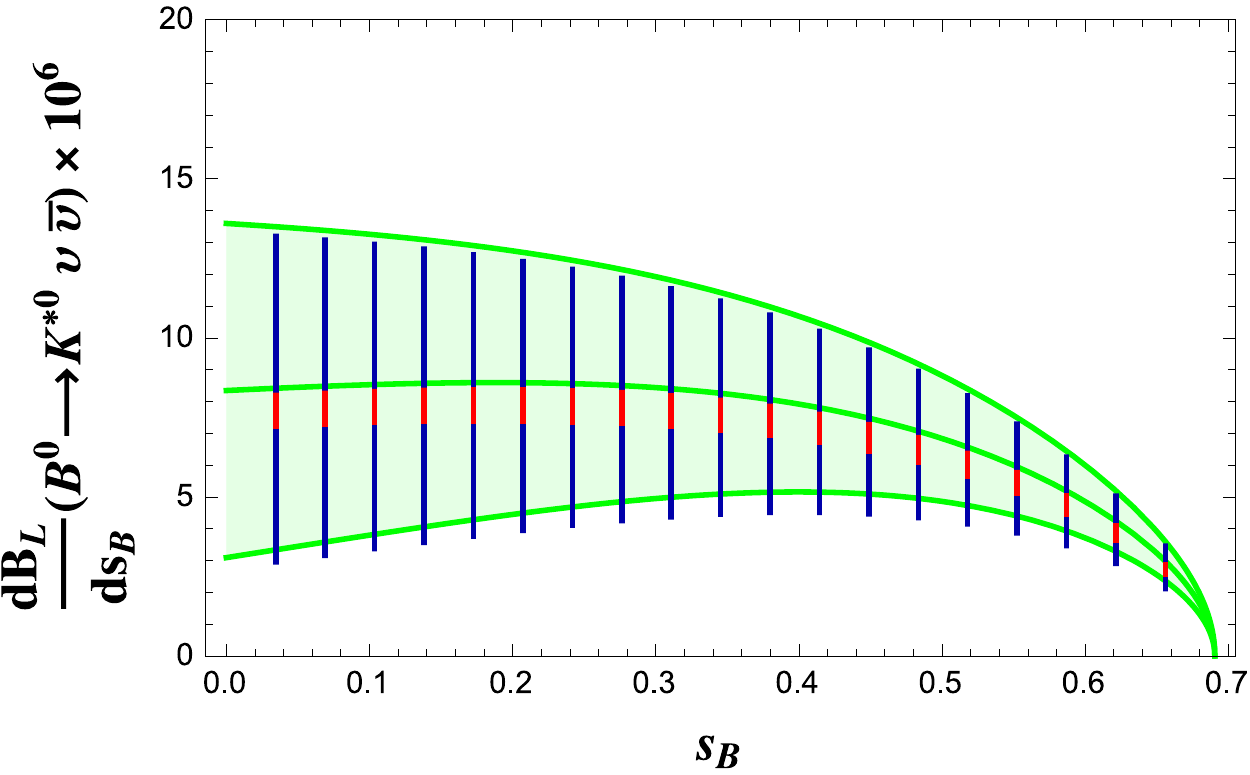}\\
\includegraphics[width = 0.4\textwidth]{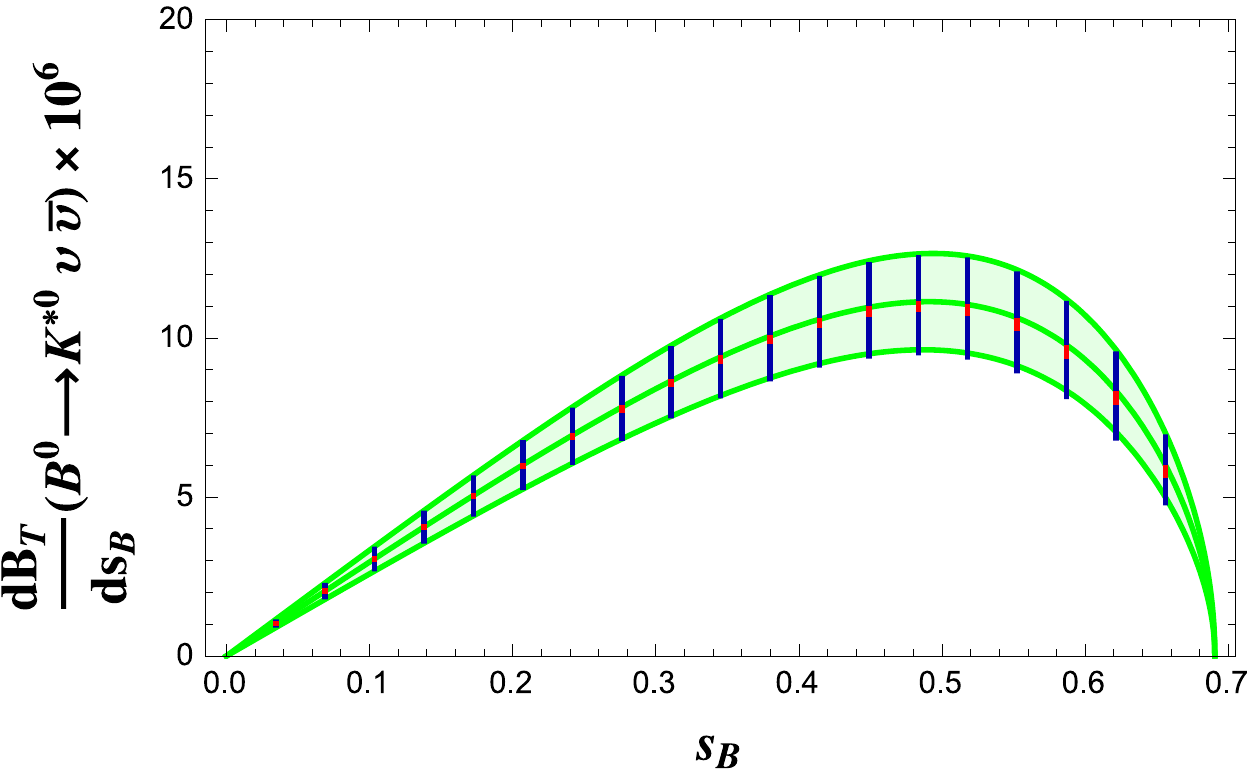}
\caption{ Distributions $\displaystyle   \frac{d {\cal B}_L }{ds_B}(B^0 \to K^{*0} \nu \bar \nu)$  (top) and $\displaystyle \frac{d {\cal B}_T }{ds_B}(B^0 \to K^{*0} \nu \bar \nu)$ (bottom).
The  green region corresponds to  SM  including the uncertainties on the form factors $A_1(0)$, $A_2(0)$ (top)  and $A_1(0)$, $V(0)$ (bottom).  The red dots and the blue bars correspond to  RS$_c$,  for the central value of the form factors and including their uncertainty at $s_B=0$, respectively.}\label{fig:dGTKstardsB}
\end{figure}
For  $B \to K^* \nu \bar \nu$   we  separately  consider the longitudinally and  transversely polarized $K^*$, with  distributions   in Fig.~\ref{fig:dGTKstardsB}.  In RS$_c$ a small  deviation from SM is found  in the longitudinal distribution.    The SM prediction, obtained including the errors on the form factors in quadrature,
\be
{\cal B}(B^0 \to K^{*0} \nu \bar \nu)_{SM}=\,(10.0 \pm2.7) \times 10^{-6} \,\,\label{BRKstarSMBall}
\ee
becomes in RS$_c$ the range
\be
{\cal B}(B^0 \to K^{*0} \nu \bar \nu)_{RS}\in[6.1 - 14.3] \times 10^{-6} \,\,.\label{BRKstarRSBall}
\ee
 For the charged mode the predictions are similar. Hence, the required improvement of the current upper bound to reach the expected signal  is about a factor of 4 in the case of
 the Belle upper bounds  (\ref{belle}), and about  one order of magnitude in the case of  the BaBar bounds (\ref{Babar}), within the reach of new facilities. Also in the case of $K^*$ our result has a more conservative error than the one quoted in (\ref{LCSR}). The difference  is due to the choice in   \cite{Altmannshofer:2008dz}  of exploiting additional information
on the  measured radiative $B \to K^* \gamma$ decay rate,  which results in a reduction of the central value and of the error of the form factors.

\begin{figure}
\includegraphics[width = 0.4\textwidth]{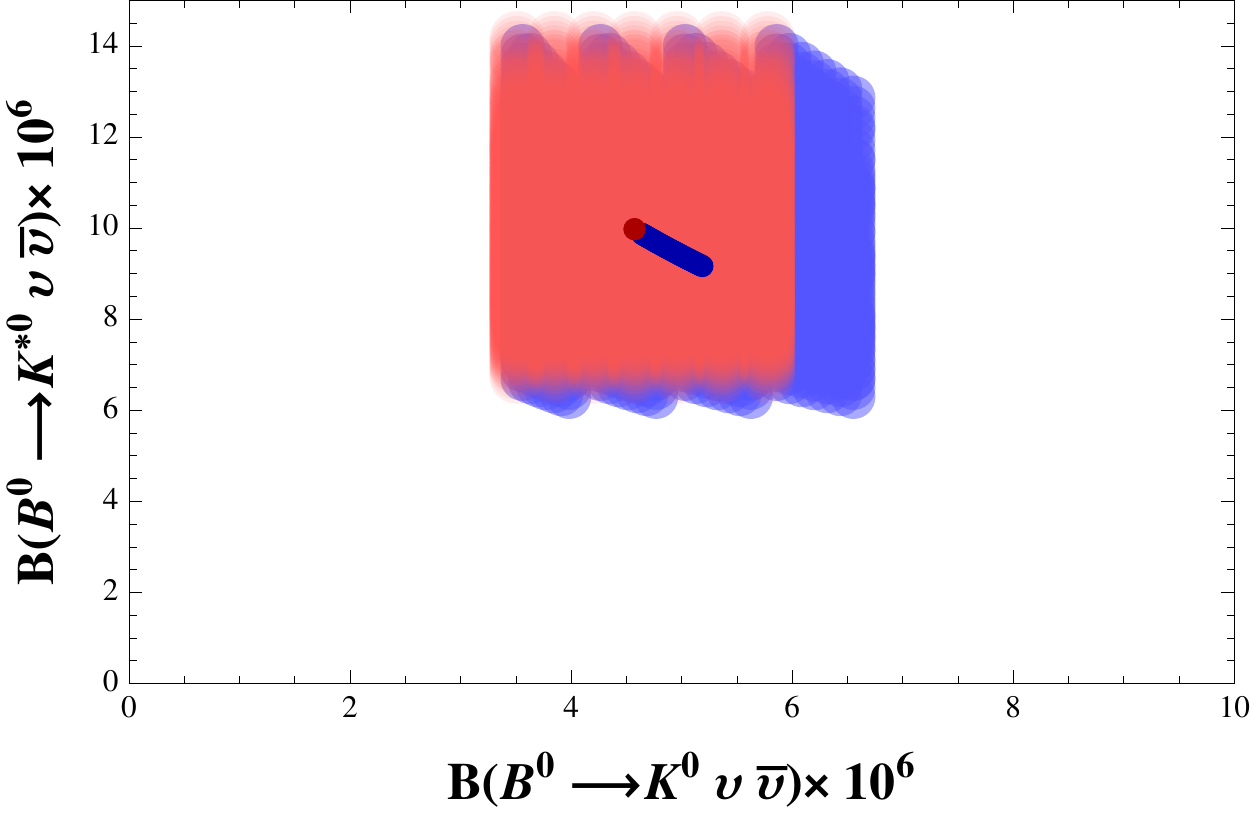}
\caption{Correlation between ${\cal B}(B^0 \to K^0 \nu \bar \nu)$  and  ${\cal B}(B^0 \to K^{*0} \nu \bar \nu)$ obtained varying the RS$_c$  parameters and including the uncertainty on  the form factors at $s_B=0$  (lighter blue region). The SM prediction corresponds to the lighter red region. The darker blue curve and the darker red dot correspond to the RS$_c$ and   SM prediction, respectively,  obtained for the central value of the form factors at $s_B=0$.  }\label{fig:BRs}
\end{figure}

\begin{figure}
\includegraphics[width = 0.4\textwidth]{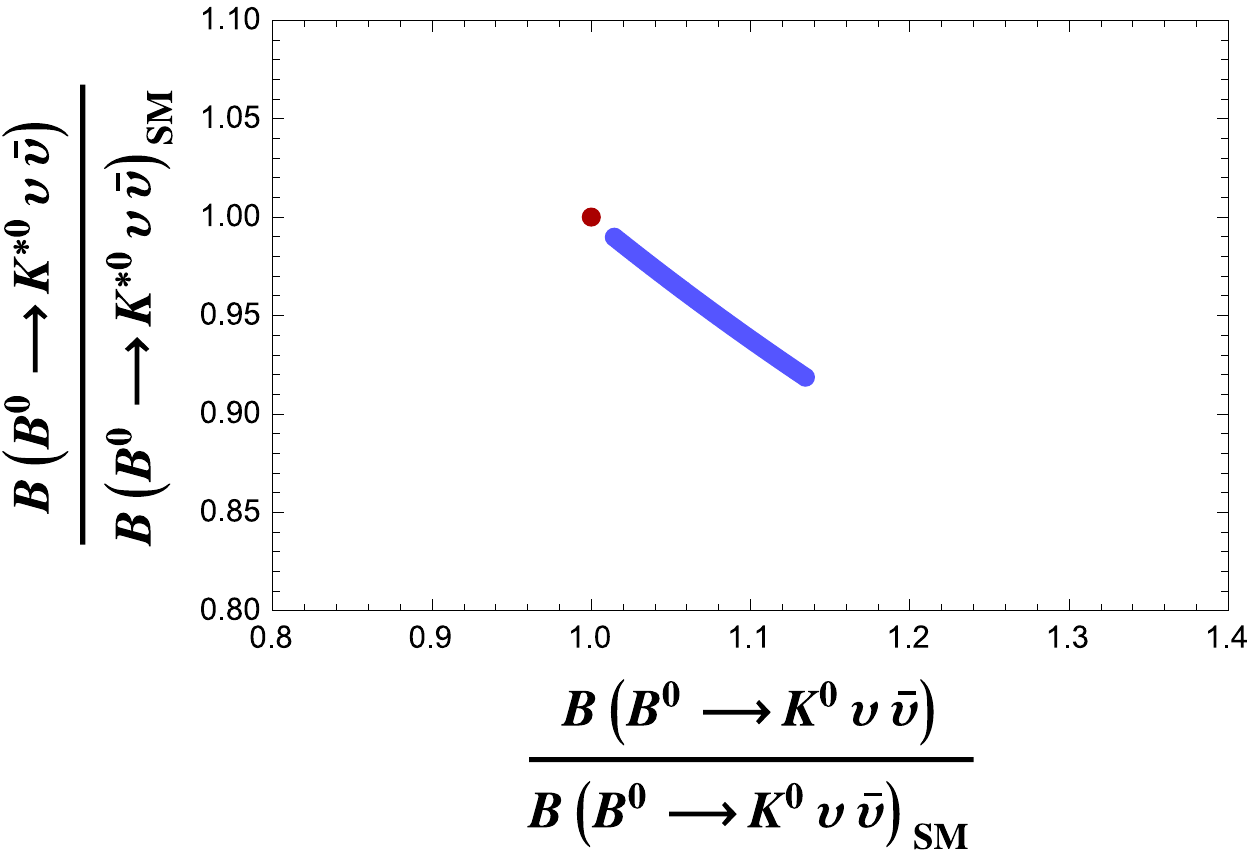}
\caption{Correlation between ${\cal B}(B^0 \to K^0 \nu \bar \nu)$  and  ${\cal B}(B^0 \to K^{*0} \nu \bar \nu)$ (blue curve) normalized to the corresponding SM values (red dot) obtained for  the central value of the form factors. }\label{fig:BRsnorm}
\end{figure}

Differently from the mode into the  pseudoscalar $K$,  the $K^*$ channel allows to access other observables as  the polarization fractions $F_{L,T}$ in    (\ref{dFLT}). Moreover, the  measurements of both the $K$ and $K^*$ modes permit the construction of the fraction $R_{K/K^*}$ in (\ref{RK-KstarT}), and  to study the correlations among the various observables predicted in SM and in RS$_c$. Such correlations are important to disentangle different NP scenarios from the one we are investigating.
In Fig.~\ref{fig:BRs}   we show the correlation between the rates of  $B \to K \nu \bar \nu$  and $B \to K^{*} \nu \bar \nu$,   with the inclusion of the hadronic uncertainty.
 Although the effects of the form factor errors are at present noticeable, the SM and the RS$_c$ predictions already have a non-overlapping region, which is interesting in view of the
 envisaged possibility of  reducing the hadronic uncertainty. In particular, the $K$ and $K^*$ modes are anticorrelated, hence a reduction of the $B \to K^* \nu \bar \nu$ decay rate goes in RS$_c$   with an increase of the rate of $B \to K \nu \bar \nu$ with respect to SM, as it is visible in Fig.~\ref{fig:BRsnorm} in the ideal case of an exact knowledge of the hadronic matrix elements.


\begin{figure}[b]
\includegraphics[width = 0.23\textwidth]{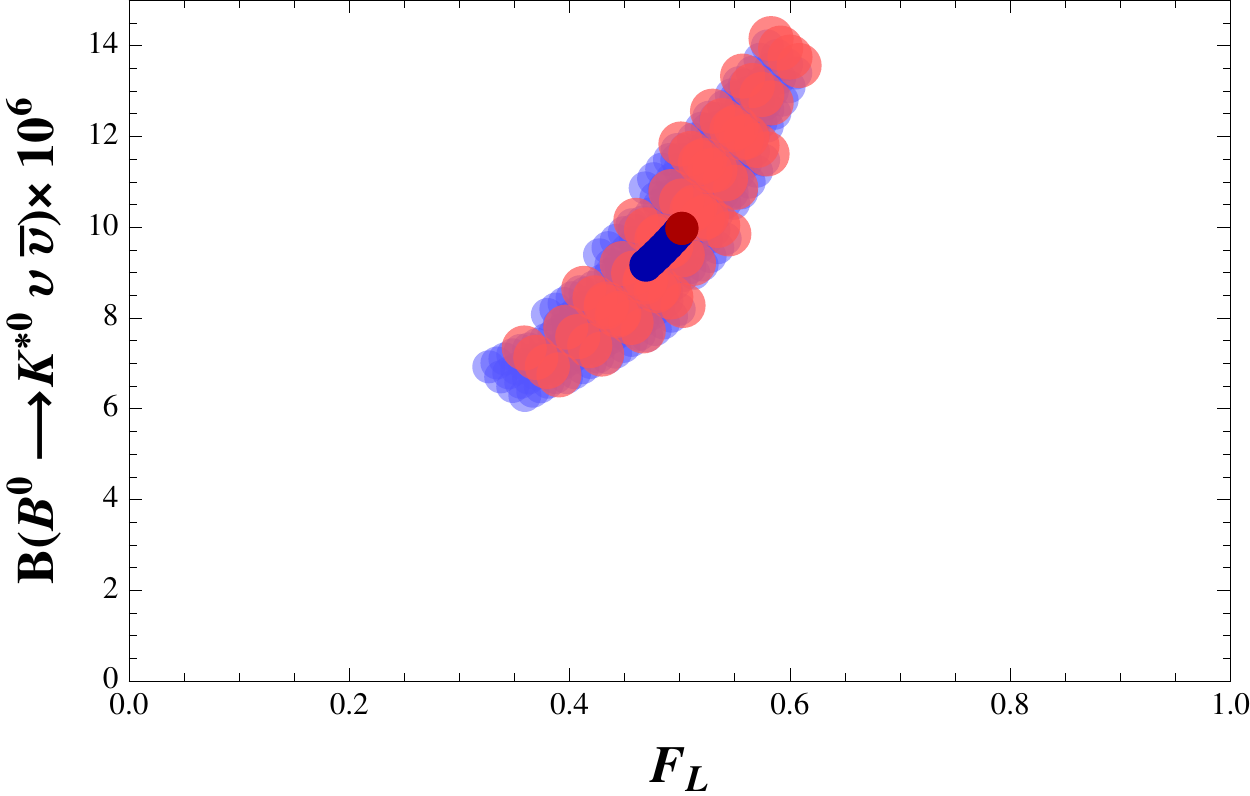}\,\, \includegraphics[width = 0.23\textwidth]{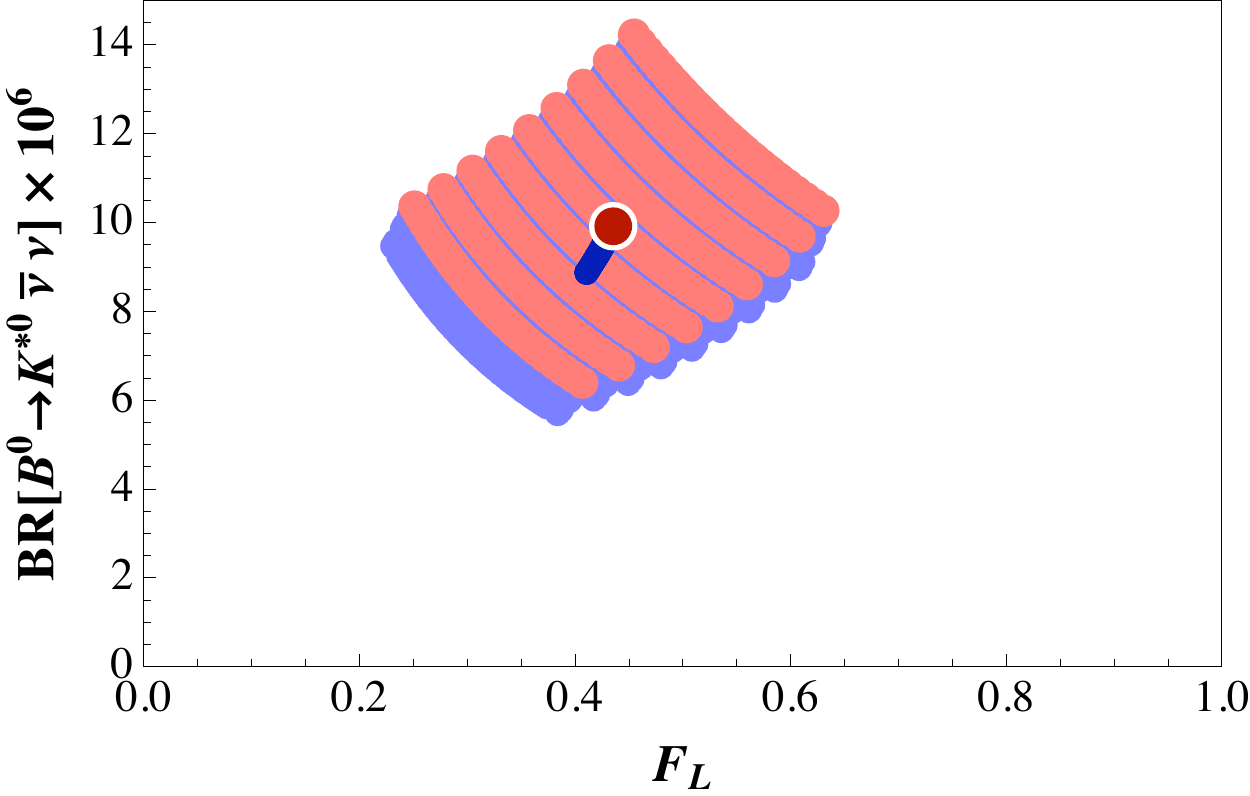}\\
\includegraphics[width = 0.23\textwidth]{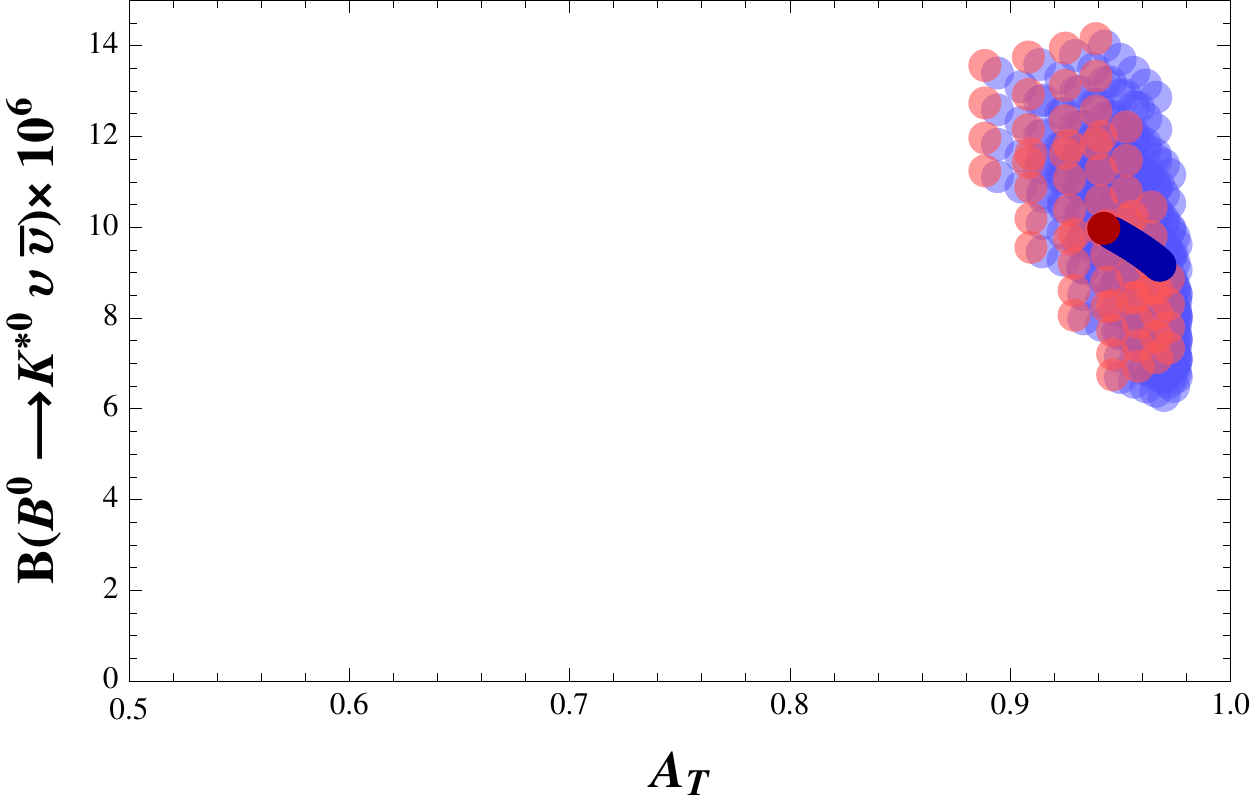}\,\, \includegraphics[width = 0.23\textwidth]{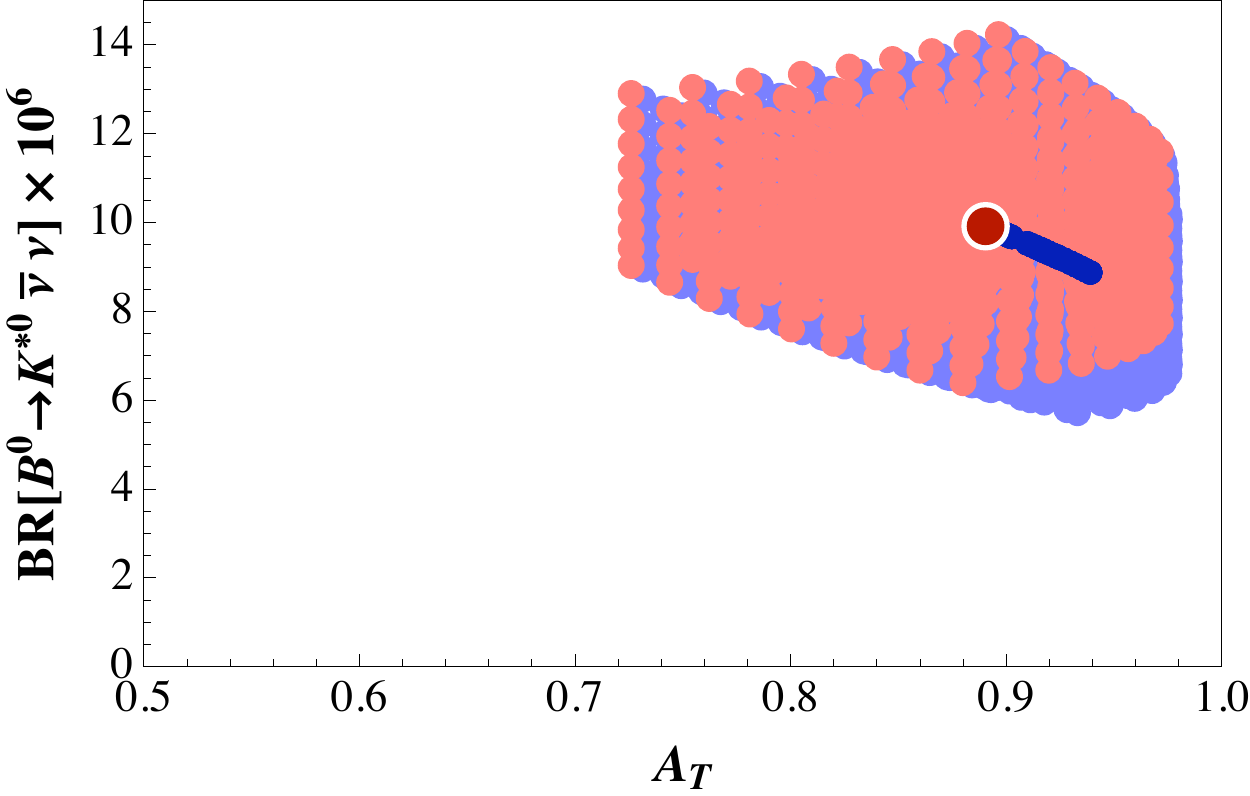}
\caption{${\cal B}(B^0 \to K^{*0} \nu \bar \nu)$ versus $F_L(B^0 \to K^{*0} \nu \bar \nu)$    (top) and   $A_T(B^0 \to K^{*0} \nu \bar \nu)$    (bottom),   varying the RS$_c$ parameters in the allowed ranges and including the uncertainty on  the form factors at $s_B=0$  (lighter blue regions)  from LCSR (left) and from lattice QCD \cite{Horgan:2013hoa} (right). The SM predictions correspond to the lighter red regions. The darker blue curves and the darker red dots correspond to the  RS$_c$ and   SM predictions, respectively,  for  the central value of the form factors. }\label{fig:AT}
\end{figure}
\begin{figure}[t!]
\includegraphics[width = 0.4\textwidth]{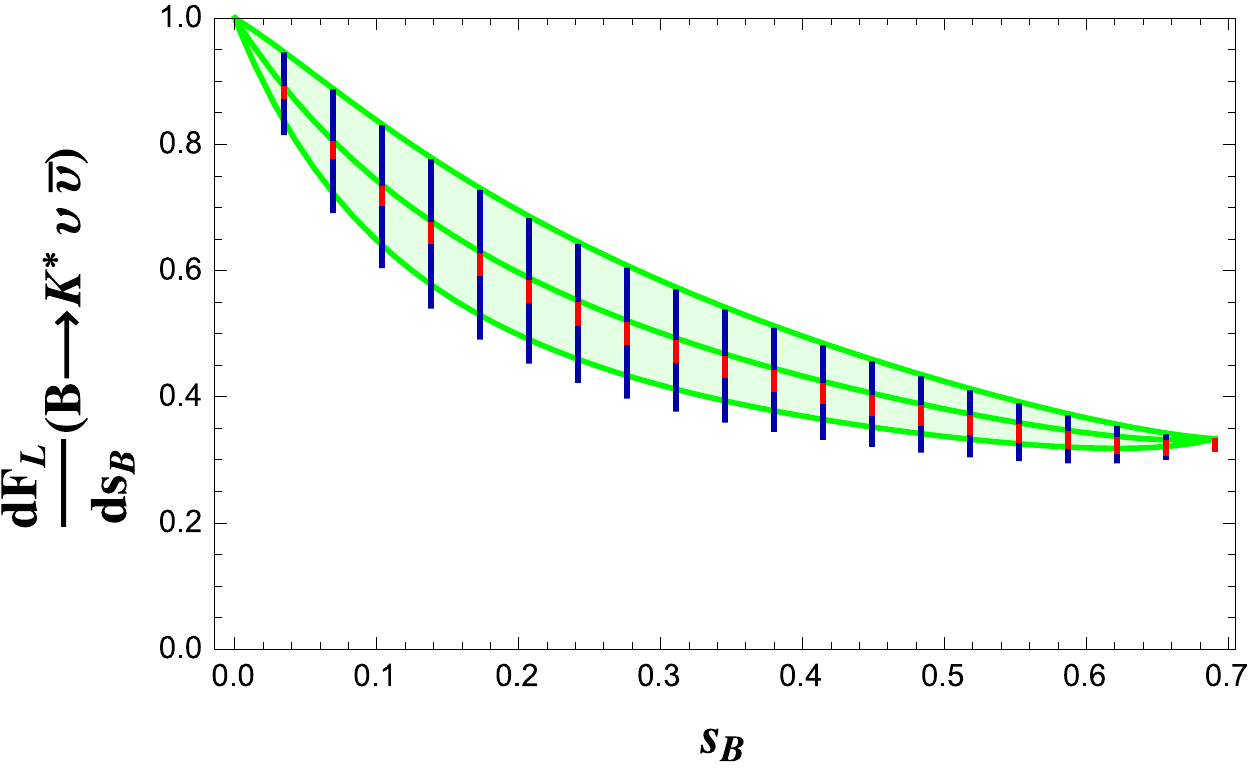}
\caption{Differential longitudinal $K^*$ polarization fraction  $\frac{d F_L}{ds_B}(B^0 \to K^{*0} \nu \bar \nu)$ in  SM including the uncertainties on  $A_1(0)$, $A_2(0)$ and $V(0)$ (green region),  and in RS$_c$ for the central values of $A_1(0)$, $A_2(0)$ and $V(0)$  (red points) and  with the error on the form factors (blue bars).  }\label{fig:dFLKstardsB}
\end{figure}
As for the longitudinal $K^*$ polarization fraction,  the differential distribution   in Fig.~\ref{fig:dFLKstardsB} has a small deviation and can be below the SM; the correlation of the integrated fraction with the branching rate is depicted in Fig.~\ref{fig:AT}.
A  precise  correlation pattern hence exists in RS$_c$ among the three observables ${\cal B}(B \to K \nu \bar \nu)$,  ${\cal B}(B \to K^* \nu \bar \nu)$ and $F_L$:  the first one can be above, the other one below  its SM values.
We also show illustratively the correlation  between  the transverse asymmetry $A_T$ in (\ref{asymT})  in  $B \to K^* \nu \bar \nu$ and the branching fraction in SM and RS$_c$,   Fig.~\ref{fig:AT}, for which  the two models, with the present hadronic uncertainty, have a big overlap.
\begin{figure}[b!]
\includegraphics[width = 0.4\textwidth]{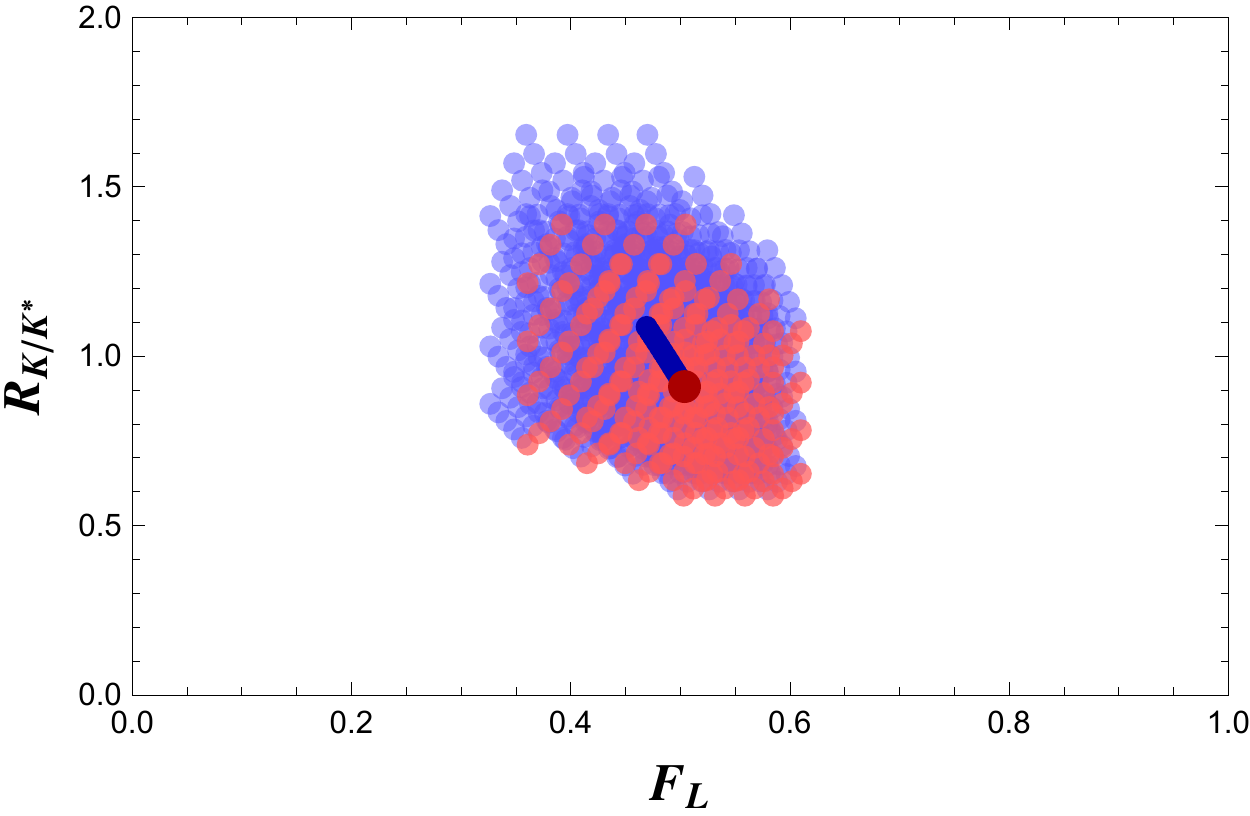}
\caption{$R_{K/K^*}$, defined in Eq.~(\ref{RK-KstarT}), versus $F_L(B^0 \to K^{*0} \nu \bar \nu)$.  The color code is the same as in Fig.~\ref{fig:AT}.}\label{fig:RKKS}
\end{figure}

The observable $R_{K/K^*}$ defined in Eq.~(\ref{RK-KstarT}) and obtained from the $K$ and $K^*$ measurements is depicted  in Fig.~\ref{fig:RKKS} versus $F_L$. A sizable form factor uncertainty is still present, at odds with the expectation that such a variable should be  quite safe; nevertheless, a region where SM and RS$_c$ results do not overlap can be observed, together with the anticorrelation with $F_L$.

An important issue concerns the hadronic error, the reliability of which cannot be  asserted without the comparison among form factors obtained by independent nonperturbative methods.
Recent lattice QCD determinations of the $B\to K^{(*)}$ form factors  \cite{Bouchard:2013eph,Horgan:2013hoa}  can be used to  estimate the size of the hadronic uncertainties affecting the various observables we have considered. Using the set in \cite{Horgan:2013hoa} we have analyzed, e.g.,  the correlation among
the $B\to K^* \nu \bar \nu$ decay rate and  $F_L$ and $A_T$. The results reported in Fig.~\ref{fig:AT}  show that the predictions already obtained  are robust  within the quoted errors.

\section{Role of the  right-handed operators  in RS$_c$}\label{sect:RHcurrents}

The correlation  between $\epsilon$ and $\eta$ is of particular interest, in light of  general analyses  where the effects of   $Z^\prime$ neutral gauge bosons are considered with no reference to  the underlying NP theory  \cite{Buras:2012jb}. In such analyses,
several possible  non-diagonal couplings to left- and right-handed quarks lead to models that can be distinguished by the relative weight of the  couplings. As an example,  a left-right symmetric scenario (LRS) corresponds to $Z^\prime$   left- and right-handed couplings equal in size and sign;   the  $\epsilon-\eta$ correlation  is different in the various cases.

Comparing our result in  Fig.~\ref{fig:etaeps} with the various possibilities considered in the general analysis,   Fig.~20 of \cite{Buras:2012jb}, we infer that the
RS$_c$ model looks similar to the RHS scenario, with  a $Z^\prime$ mainly coupled  to right-handed quarks.  Indeed,  the difference $C_L^{RS}-C_L^{SM}$ and the coefficient $C_R^{RS}$,  playing the role of the left- and right-handed quark couplings to a  new gauge boson, have opposite  sign,   and  $C_R^{RS} \ll C_L^{RS}-C_L^{SM}$, Fig.~\ref{fig:CLR}.
Although in  RS$_c$ there are several additional gauge boson,  the  effect is similar  to the case of one new boson.

\begin{figure}[b]
\includegraphics[width = 0.4\textwidth]{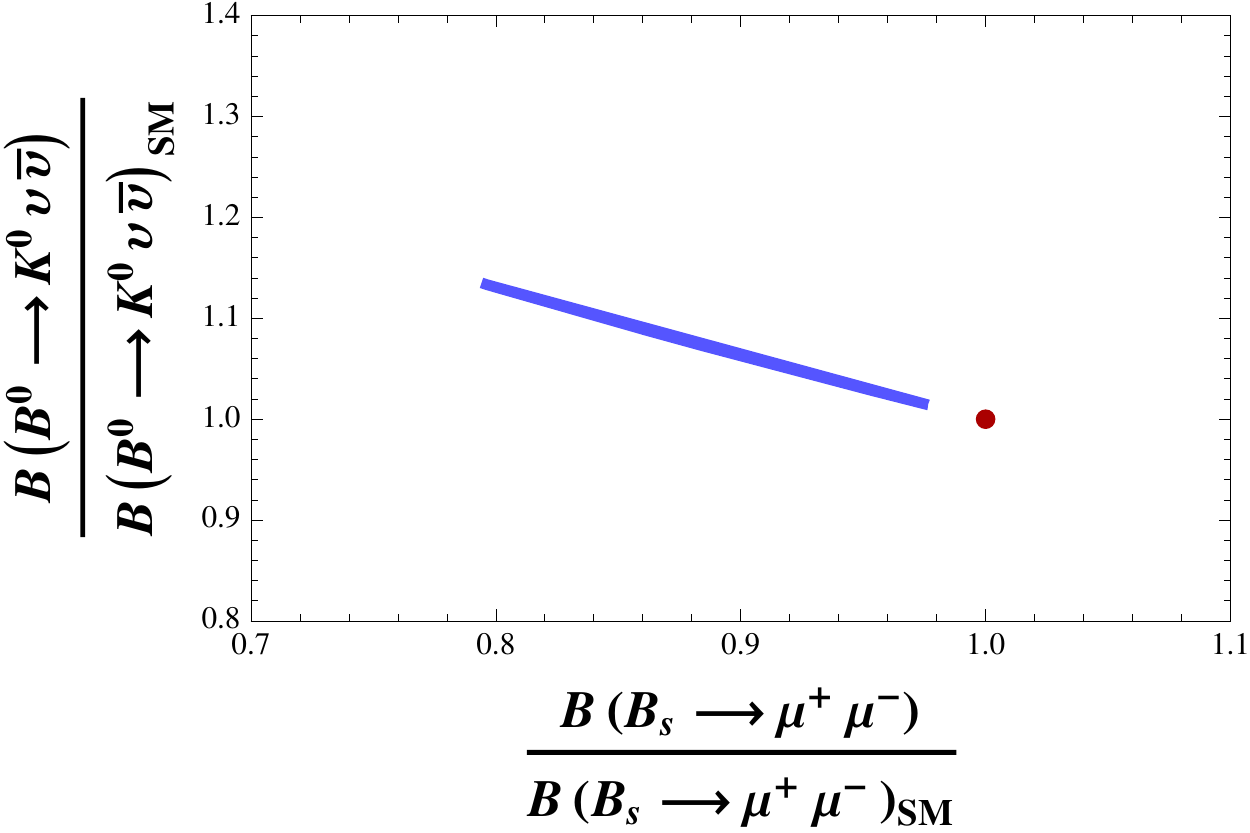}\\
\includegraphics[width = 0.4\textwidth]{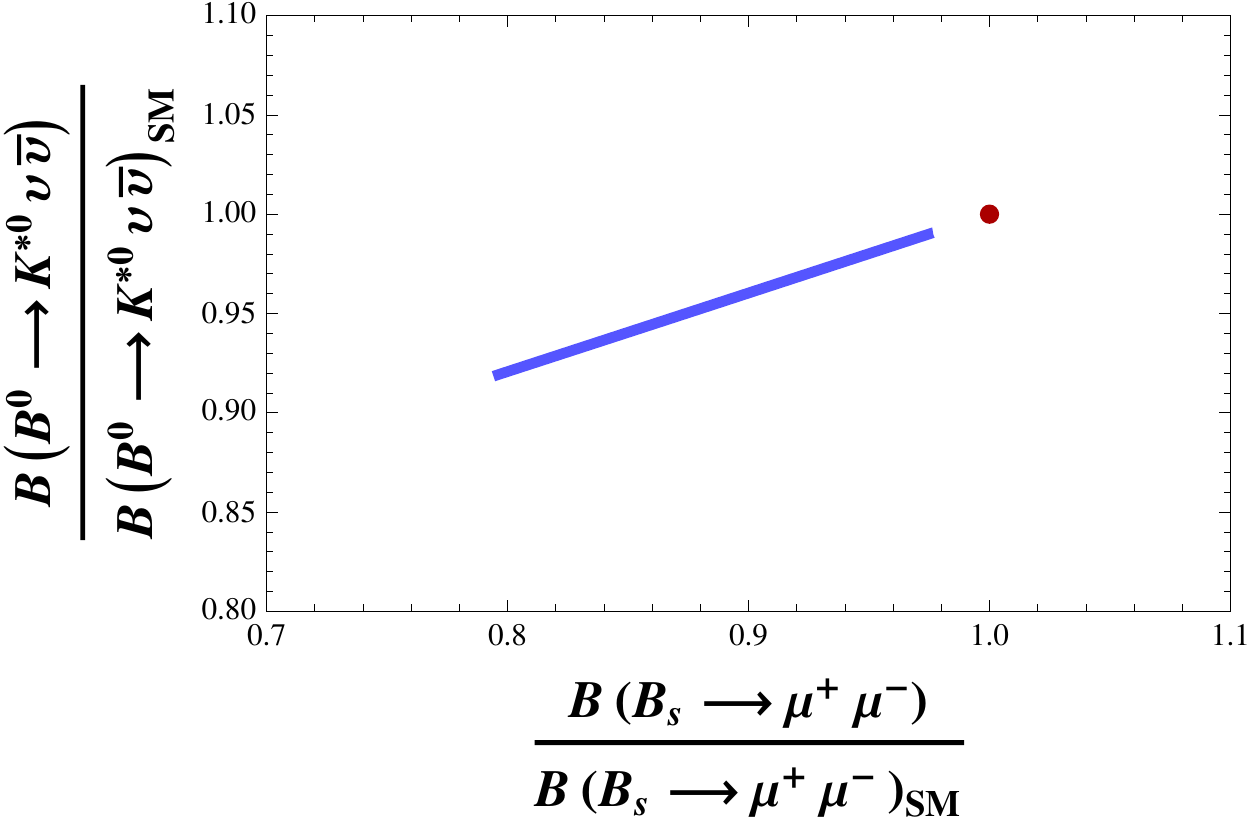}
\caption{Correlation between  ${\cal B}(B^0 \to K^0 \nu \bar \nu)$  and  ${\cal B}(B_s \to \mu \bar \mu)$ (top), and between ${\cal B}(B^0 \to K^{*0} \nu \bar \nu)$  and  ${\cal B}(B_s  \to \mu \bar \mu)$ (bottom)   normalized to their central SM values. The hadronic uncertainty is not included.  The blue lines correspond to RS$_c$, the red dots to SM.}\label{fig:BRKstarBs}
\end{figure}

The correlation of  ${\cal B}(B \to K \nu \bar \nu)$ and  ${\cal B}(B \to K^* \nu \bar \nu)$ with ${\cal B}(B_s \to \mu^+ \mu^-)$ provides  a deeper insight.
In NP models one has
\be
\frac{{\cal B}(B_s \to \mu^+ \mu^-)}{{\cal B}(B_s \to \mu^+ \mu^-)|_{SM}}=\frac{C_{10}-C_{10}^\prime}{C_{10}^{SM}}
\ee
where $C_{10}$ and  $C_{10}^\prime$ are the Wilson coefficients of the semileptonic electroweak penguin operators with axial vector  leptonic current  and $V-A$ and $V+A$ structure of the quark current in the effective $b \to s \ell^+ \ell^-$ Hamiltonian.   In  SM   only $C_{10}^{SM}$ is relevant  (the contribution of  $O^\prime_{10}$ is negligible).
Evaluating  $C_{L,R}$, $C_{10}$ and $C_{10}^\prime$ in  the RS$_c$  parameter space, the  correlations in Fig.~\ref{fig:BRKstarBs} are found.   The rates of $B \to K \nu \bar \nu$ and $B_s \to \mu^+ \mu^-$    are anticorrelated:  in  RS$_c$ a larger ${\cal B}(B_s \to \mu^+ \mu^-)$ than in  SM   implies a   lower  ${\cal B}(B \to K \nu \bar \nu)$. The opposite happens for $B \to K^* \nu \bar \nu$:  ${\cal B}(B \to K^* \nu \bar \nu)$ and ${\cal B}(B_s \to \mu^+ \mu^-)$ are correlated,   therefore finding one of them above its SM value would require an enhancement also of the other one.
This again characterizes RS$_c$  as  an RHS scenario, as one can infer from a comparison with the general result  of Fig.~21 in \cite{Buras:2012jb}.

\section{$B_s \to (\phi, \eta, \eta^\prime, f_0(980)) \bar \nu \nu$ in RS$_c$}\label{sect:Bs}
Several $B_s$ decay modes of great phenomenological interest are driven by the transition $b \to s \nu \bar \nu$.
Here we focus on  $B_s \to ( \eta, \eta^\prime)  \nu \bar \nu$,   on the decay $B_s \to \phi  \nu \bar \nu$ and  on $B_s \to  f_0(980)  \nu \bar \nu$ with the scalar  $f_0(980)$ meson in the final state, all of them  accessible at the new facilities.

The modes into  $\eta$ and $\eta^\prime$  must be considered altogether, due to the  $\eta-\eta^\prime$ mixing.
Two  schemes are usually adopted to  describe this mixing, in either  the singlet-octet (SO)  or  the quark-flavor (QF) basis, and  both schemes involve  two mixing angles  \cite{Feldmann:1999uf}.
We choose the quark-flavor basis, defining
 \bea \ket{\eta_q}&=&{1 \over \sqrt{2} } \left(\ket{{\bar u} u} +\ket{{\bar d} d}\right) \nonumber \\
\ket{\eta_s}&=& \ket{{\bar s} s} \,\, , \label{etaqs}
\eea
in which the two  mixing angles  $\varphi_q$ and $\varphi_s$,
\bea
\ket{\eta}&=&  \cos \, \varphi_q \ket{\eta_q}- {\sin} \, \varphi_s  \ket{\eta_s} \nonumber \\
\ket{\eta^\prime}&=& { \sin} \, \varphi_q  \ket{\eta_q}+{\cos} \, \varphi_s \ket{\eta_s}  \,\,\,\, , \label{mixing}
\eea
differ by OZI-violating effects.  However, the difference is experimentally  found  to be small,
($\varphi_q-\varphi_s  <  5^\circ$),  therefore,  within the present accuracy we can adopt an $\eta-\eta^\prime$ mixing description  in the QF basis and a single mixing angle   $\varphi_q\simeq \varphi_s \simeq \varphi$.  This choice is supported  by a  study
of the  radiative  $\phi \to \eta\gamma$ and  $\phi \to \eta^\prime\gamma$ transitions  \cite{DeFazio:2000my}.
The KLOE Collaboration has measured
 the ratio $\displaystyle{ \Gamma(\phi \to \eta^\prime  \gamma) \over \Gamma(\phi \to \eta \gamma)}$,  finding for the $\eta-\eta^\prime$ mixing angle  the value
$\varphi=\big( 41.5 \pm 0.3_{stat} \pm 0.7_{syst} \pm0.6_{th} \big )^\circ$   \cite{Ambrosino:2006gk}.
An improved  analysis by the same collaboration, allowing a gluonium content in the $\eta^\prime$ and  making use of the measured  ratio $\displaystyle {\Gamma (\eta^\prime \to \gamma \gamma) \over \Gamma(\pi^0 \to \gamma \gamma)}$   confirms this  determination of  $\varphi$ \cite{Ambrosino:2009sc}.

The flavour symmetry permits to relate   the  $B_s \to \eta,\,\eta^\prime$ form factors  to the $B \to K$ ones.  For a  form factor $F$ one has  $F^{B_s \to \eta}=-\sin \varphi F^{B \to K}$ and $F^{B_s \to \eta^\prime}=\cos \varphi F^{B \to K}$
 \cite{Carlucci:2009gr}. On the other hand, for the $B_s \to \phi \nu \bar \nu$ mode we use  the LCSR  $B_s \to \phi$ form factors in  Ref.~\cite{Ball:2004rg}.

\begin{figure}
\includegraphics[width = 0.4\textwidth]{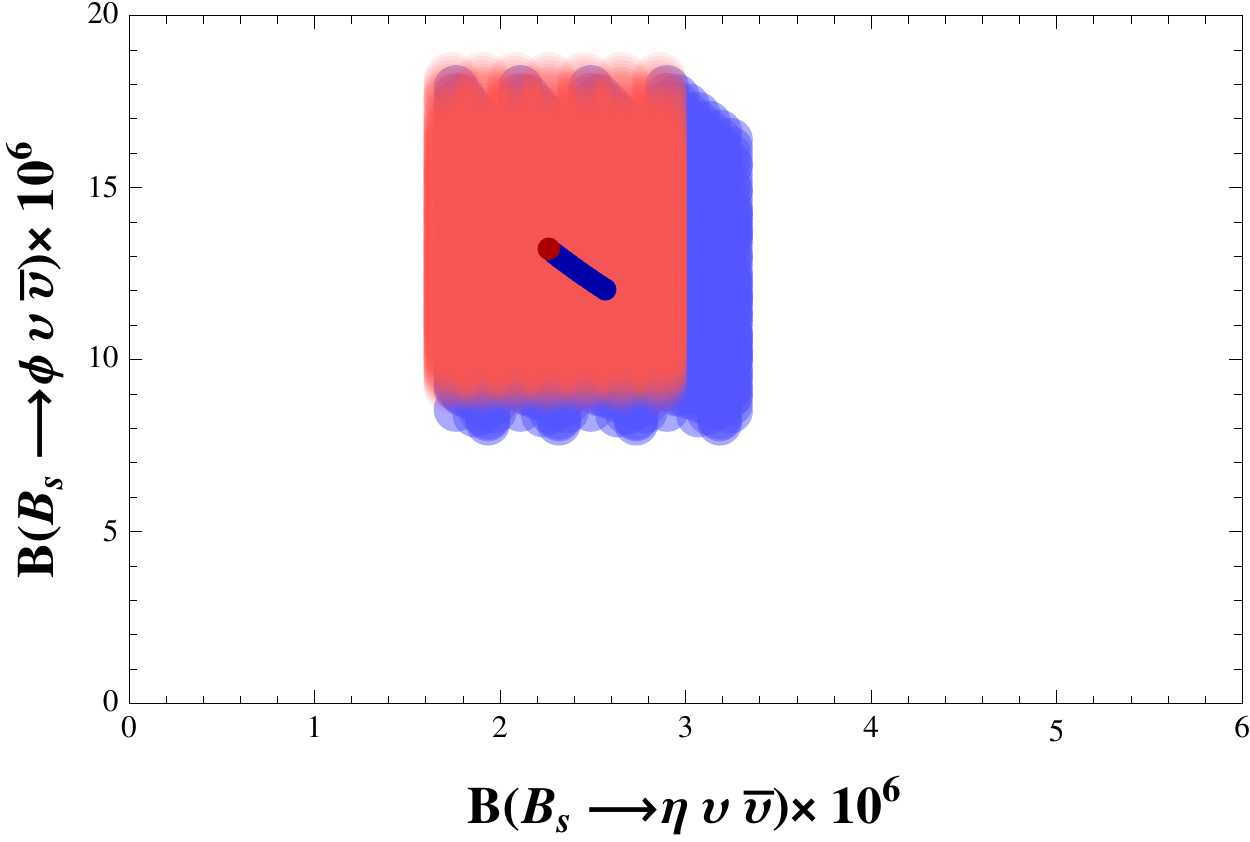}\\
\includegraphics[width = 0.4\textwidth]{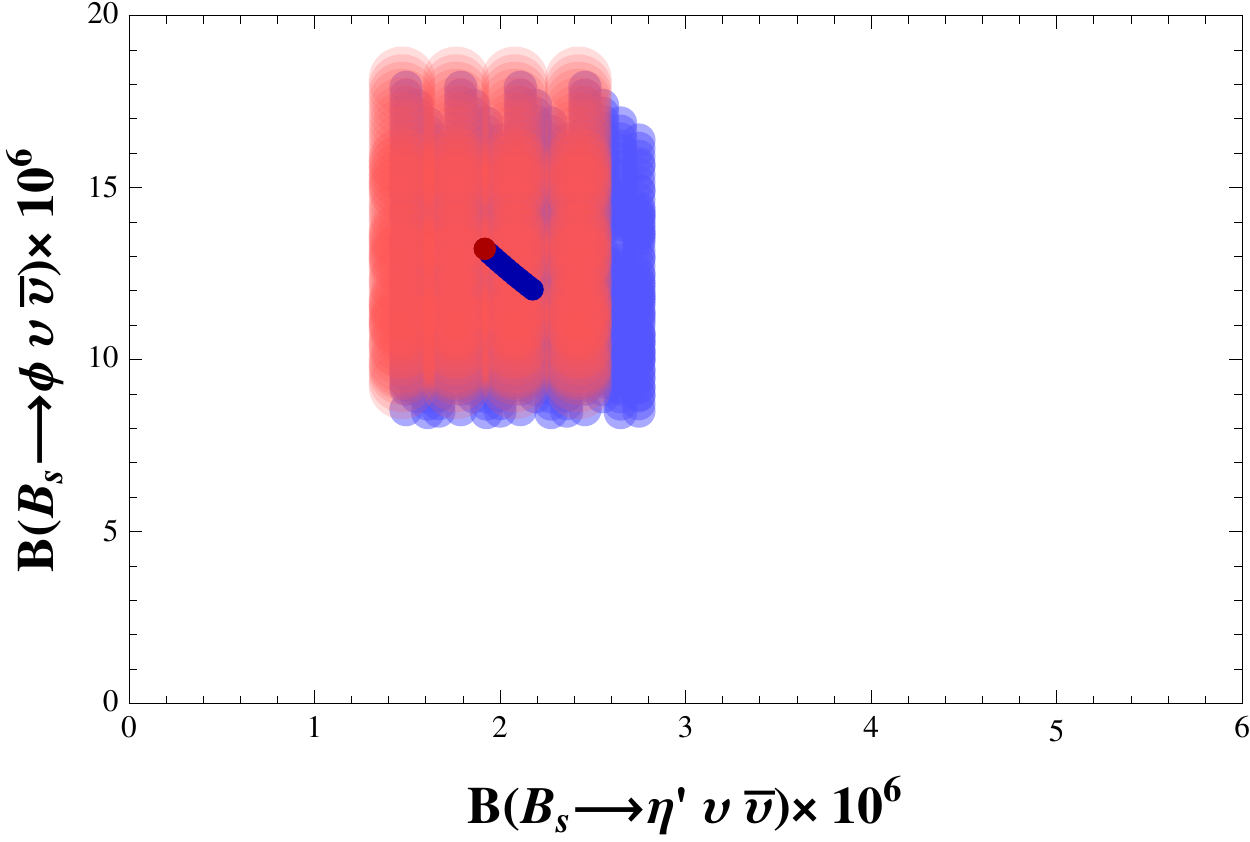}\\
\includegraphics[width = 0.4\textwidth]{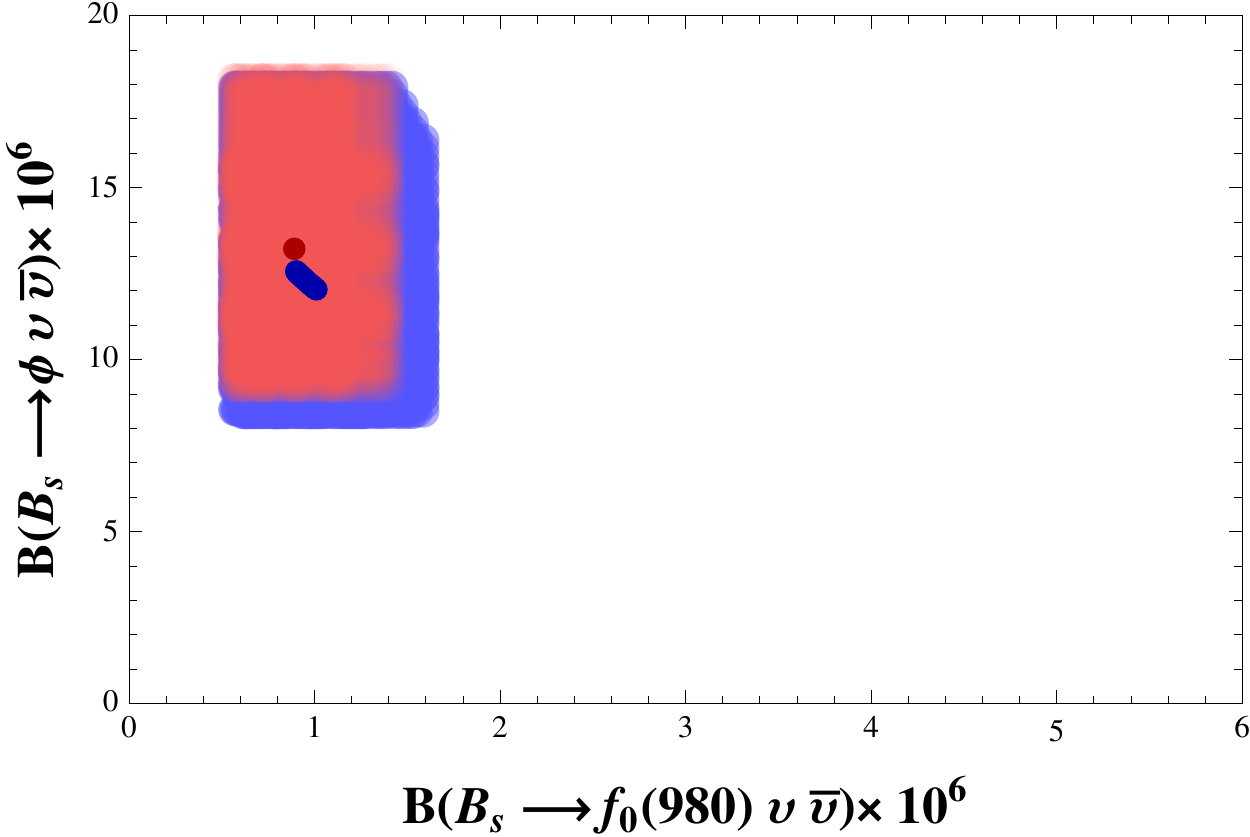}
\caption{Correlation of ${\cal B}(B_s \to \phi  \nu \bar \nu)$ with ${\cal B}(B_s \to  \eta  \nu \bar \nu)$ (top),  ${\cal B}(B_s \to  \eta^\prime  \nu \bar \nu)$ (center) and ${\cal B}(B_s \to  f_0(980)  \nu \bar \nu)$ (bottom). The color code is the same as in Fig.\ref{fig:AT}.}\label{fig:BRetaFF}
\end{figure}

\begin{figure}[b!]
\includegraphics[width = 0.4\textwidth]{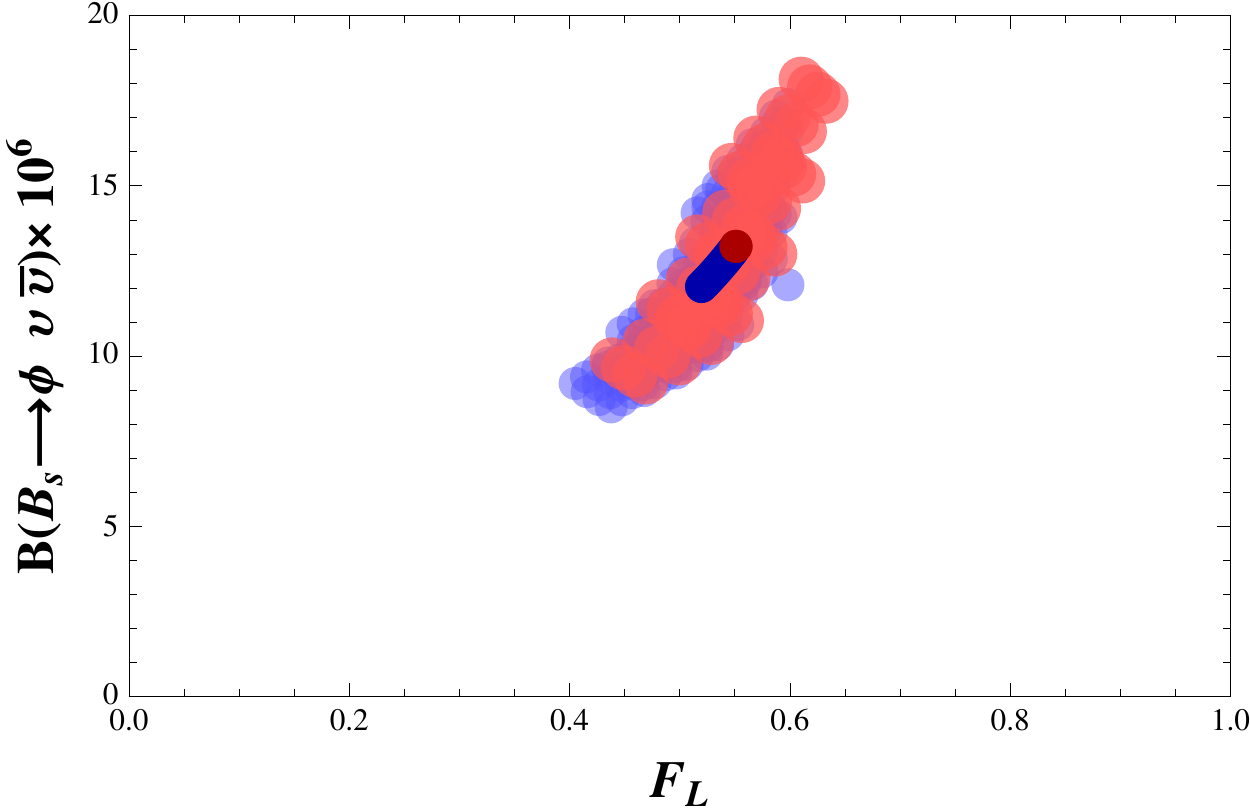}
\caption{ ${\cal B}(B_s \to \phi  \nu \bar \nu)$ versus $F_L(B_s \to  \phi  \nu \bar \nu)$. The color code is the same as in Fig.\ref{fig:AT}.}\label{fig:phiFLAT}
\end{figure}

The SM predictions, obtained for $\tau (B_s)=1.512\pm0.007$~ps \cite{Amhis:2012bh},
\bea
{\cal B}(B_s \to \eta \nu \bar \nu)_{SM}&=&\,(2.3 \pm 0.5) \times 10^{-6} \,\,\label{BRetaSMBall}\\
{\cal B}(B_s \to \eta^\prime \nu \bar \nu)_{SM}&=&\,(1.9 \pm 0.5) \times 10^{-6} \,\,\label{BRetapSMBall}\\
{\cal B}(B_s \to \phi \nu \bar \nu)_{SM}&=&\,(13.2 \pm3.3) \times 10^{-6} \,\,\label{BRphiSMBall}
\eea
are modified in RS$_c$:
\bea
{\cal B}(B_s \to \eta \nu \bar \nu)_{RS}&&\in[1.7 - 3.3] \times 10^{-6} \,\,\label{BRetaRSBall}\\
{\cal B}(B_s \to \eta^\prime \nu \bar \nu)_{RS}&&\in[1.5 - 2.8] \times 10^{-6} \,\,\label{BRetapRSBall}\\
{\cal B}(B_s \to \phi \nu \bar \nu)_{RS}&&\in[8.4 - 18.0] \times 10^{-6} \,\,.\label{BRphiRSBall}
\eea
The result is particularly relevant in the case of the $B_s \to \phi$ mode, which should be the first one accessible for the $B_s$ meson: the rate is within the reach of the new facilities,
the $\phi$ can be easily identified and its decay modes  allow to  construct, e.g., the $F_L$ observable.
The various correlation patterns are shown in Fig.~\ref{fig:BRetaFF}:  anticorrelation
is found between the rates of $B_s \to \eta^{(\prime)} \nu \bar \nu$  with $B_s \to \phi \nu \bar \nu$. For  $F_L$ the results are depicted in  Fig.~\ref{fig:phiFLAT}.

The last mode in our analysis involves the scalar $f_0(980)$ meson.
The $B_s \to f_0(980)$ form factors have been determined assuming for  $f_0(980)$ a dominant  quark-antiquark $s {\bar s}$ structure \cite{Colangelo:2010bg}.  With the updated value of $\tau(B_s)$  we find that the SM prediction is modified in RS$_c$:
\bea
{\cal B}(B_s \to f_0(980) \nu \bar \nu)_{SM}&=&(8.95 \pm^{2.9}_{2.5})\times 10^{-7}  \label{BRf0} \\
{\cal B}(B_s \to f_0(980) \nu \bar \nu)_{RS}&\in&[5 - 17] \times 10^{-7} \,\, .\label{BRf0RSnoi}
\eea
This channel should be  accessible through the  $f_0(980)\to \pi \pi$ transition,  providing  another test of the RS$_c$ scenario.

\section{ Conclusions}\label{sec:conclusions}
Although experimentally challenging, the FCNC exclusive $b$-hadron  transitions into $\nu \bar \nu$ pairs are of great interest, as they can provide the evidence of possible deviations from SM
through signals of remarkable theoretical significance. We have exhamined a set of $B$ and $B_s$ decay modes in the RS$_c$ model, with particular emphasis on the correlations among the observables  that are  features of the model. In the planned experimental analyses these modes can be accessible, and the  predictions  presented here will become testable.



\begin{thebibliography}{99}

\bibitem{general-review}
G.~Isidori, Y.~Nir and G.~Perez,
Ann.\ Rev.\ Nucl.\ Part.\ Sci.\  {\bf 60}, 355 (2010)
[\href{http://arxiv.org/abs/arXiv:1002.0900}{arXiv:1002.0900 [hep-ph]}];
A.~J.~Buras and J.~Girrbach,
\href{http://arxiv.org/abs/arXiv:1306.3775}{arXiv:1306.3775  [hep-ph]};
G.~Isidori and F.~Teubert,
Eur.\ Phys.\ J.\ Plus {\bf 129}, 40 (2014)
[\href{http://arxiv.org/abs/arXiv:1402.2844}{arXiv:1402.2844 [hep-ph]}].

\bibitem{Buras:1998raa}
  A.~J.~Buras,
  \href{http://arxiv.org/abs/hep-ph/9806471}{hep-ph/9806471}.

\bibitem{Colangelo:1996ay}
  P.~Colangelo, F.~De Fazio, P.~Santorelli and E.~Scrimieri,
  Phys.\ Lett.\ B {\bf 395}, 339 (1997)
  [\href{http://arxiv.org/abs/hep-ph/9610297}{hep-ph/9610297}].

\bibitem{Melikhov:1998ug}
  D.~Melikhov, N.~Nikitin and S.~Simula,
  Phys.\ Lett.\ B {\bf 428}, 171 (1998)
    [\href{http://arxiv.org/abs/hep-ph/9803269}{hep-ph/9803269}].

\bibitem{Altmannshofer:2008dz}
W.~Altmannshofer, P.~Ball, A.~Bharucha, A.~J.~Buras, D.~M.~Straub and M.~Wick,
JHEP {\bf 0901}, 019 (2009)
[\href{http://arxiv.org/abs/arXiv:0811.1214}{arXiv:0811.1214  [hep-ph]}].

\bibitem{Altmannshofer:2009ma}
  W.~Altmannshofer, A.~J.~Buras, D.~M.~Straub and M.~Wick,
  JHEP {\bf 0904}, 022 (2009)
  [\href{http://arxiv.org/abs/arXiv:0902.0160}{arXiv:0902.0160  [hep-ph]}].

\bibitem{Lutz:2013ftz}
  O.~Lutz {\it et al.}  [Belle Collaboration],
  Phys.\ Rev.\ D {\bf 87}, no. 11, 111103 (2013)
    [\href{http://arxiv.org/abs/arXiv:1303.3719}{arXiv:1303.3719  [hep-ex]}].

\bibitem{Lees:2013kla}
  J.~P.~Lees {\it et al.}  [BaBar Collaboration],
  Phys.\ Rev.\ D {\bf 87}, no. 11, 112005 (2013)
        [\href{http://arxiv.org/abs/arXiv:1303.7465}{arXiv:1303.7465  [hep-ex]}].

  \bibitem{delAmoSanchez:2010bk}
  P.~del Amo Sanchez {\it et al.}  [BaBar Collaboration],
  Phys.\ Rev.\ D {\bf 82}, 112002 (2010)
      [\href{http://arxiv.org/abs/arXiv:1009.1529}{arXiv:1009.1529  [hep-ex]}].

\bibitem{Beringer:1900zz}
J.~Beringer {\it et al.}  [Particle Data Group Collaboration],
Phys.\ Rev.\ D {\bf 86}, 010001 (2012).

\bibitem{Amhis:2012bh}
Y.~Amhis {\it et al.}  [Heavy Flavor Averaging Group Collaboration],
\href{http://arxiv.org/abs/arXiv:1207.1158}{arXiv:1207.1158  [hep-ph]} and http://www.slac.stanford.edu/xorg/hfag/.

\bibitem{Kim:1999waa}
C.~S.~Kim, Y.~G.~Kim and T.~Morozumi,
Phys.\ Rev.\ D {\bf 60}, 094007 (1999)
 [\href{http://arxiv.org/abs/hep-ph/9905528}{hep-ph/9905528}].

\bibitem{Buchalla:2000sk}
  G.~Buchalla, G.~Hiller and G.~Isidori,
  Phys.\ Rev.\ D {\bf 63}, 014015 (2000)
     [\href{http://arxiv.org/abs/hep-ph/0006136}{hep-ph/0006136}].

\bibitem{Buras:2012jb}
  A.~J.~Buras, F.~De Fazio and J.~Girrbach,
  JHEP {\bf 1302}, 116 (2013)
  [\href{http://arxiv.org/abs/arXiv:1211.1896}{arXiv:1211.1896 [hep-ph]}].

\bibitem{Colangelo:2006vm}
  P.~Colangelo, F.~De Fazio, R.~Ferrandes and T.~N.~Pham,
  Phys.\ Rev.\ D {\bf 73}, 115006 (2006)
      [\href{http://arxiv.org/abs/hep-ph/0604029}{hep-ph/0604029}].

\bibitem{RandallSundrum}
L.~Randall and R.~Sundrum,
  Phys.\ Rev.\ Lett.\  {\bf 83}, 3370 (1999)
  [\href{http://arxiv.org/abs/hep-ph/9905221}{hep-ph/9905221}].

\bibitem{contino}
K.~Agashe, R.~Contino, L.~Da Rold and A.~Pomarol,
Phys.\ Lett.\ B {\bf 641}, 62 (2006)
[\href{http://arxiv.org/abs/hep-ph/0605341}{hep-ph/0605341}].

\bibitem{carena}
M.~S.~Carena, E.~Ponton, J.~Santiago and C.~E.~M.~Wagner,
Nucl.\ Phys.\ B {\bf 759}, 202 (2006)
[\href{http://arxiv.org/abs/hep-ph/0607106}{hep-ph/0607106}].

\bibitem{cacciapaglia}
G.~Cacciapaglia, C.~Csaki, G.~Marandella and J.~Terning,
Phys.\ Rev.\ D {\bf 75}, 015003 (2007)
[\href{http://arxiv.org/abs/hep-ph/0607146}{hep-ph/0607146}].

\bibitem{buras2}
M.~Blanke, A.~J.~Buras, B.~Duling, K.~Gemmler and S.~Gori,
JHEP {\bf 0903}, 108 (2009)
[\href{http://arxiv.org/abs/arXiv:0812.3803}{arXiv:0812.3803 [hep-ph]}].

\bibitem{Biancofiore:2014wpa}
  P.~Biancofiore, P.~Colangelo and F.~De Fazio,
 Phys.\ Rev.\ D {\bf 89}, 095018 (2014)
[\href{http://arxiv.org/abs/arXiv:1403.2944}{arXiv:1403.2944  [hep-ph]}].

\bibitem{Inami:1980fz}
  T.~Inami and C.~S.~Lim,
  Prog.\ Theor.\ Phys.\  {\bf 65}, 297 (1981)
  [Erratum-ibid.\  {\bf 65}, 1772 (1981)].

\bibitem{Buchalla:1998ba}
  G.~Buchalla and A.~J.~Buras,
  Nucl.\ Phys.\ B {\bf 548}, 309 (1999)
    [\href{http://arxiv.org/abs/hep-ph/9901288}{hep-ph/9901288}].

\bibitem{Falk:1995kn}
  A.~F.~Falk, M.~E.~Luke and M.~J.~Savage,
  Phys.\ Rev.\ D {\bf 53}, 6316 (1996)
  [\href{http://arxiv.org/abs/hep-ph/9511454}{hep-ph/9511454}].

\bibitem{Egede:2008uy}
  U.~Egede, T.~Hurth, J.~Matias, M.~Ramon and W.~Reece,
  JHEP {\bf 0811}, 032 (2008)
  [\href{http://arxiv.org/abs/arXiv:0807.2589}{arXiv:0807.2589  [hep-ph]}].

\bibitem{Albrecht:2009xr}
 M.~E.~Albrecht, M.~Blanke, A.~J.~Buras, B.~Duling and K.~Gemmler,
JHEP {\bf 0909}, 064 (2009)
[\href{http://arxiv.org/abs/arXiv:0903.2415}{arXiv:0903.2415  [hep-ph]}].

\bibitem{buras1}
M.~Blanke, A.~J.~Buras, B.~Duling, S.~Gori and A.~Weiler,
JHEP {\bf 0903}, 001 (2009)
[\href{http://arxiv.org/abs/arXiv:0809.1073}{arXiv:0809.1073  [hep-ph]}].

\bibitem{buras4}
A.~J.~Buras, B.~Duling and S.~Gori,
JHEP {\bf 0909}, 076 (2009)
[\href{http://arxiv.org/abs/arXiv:0905.2318}{arXiv:0905.2318  [hep-ph]}].

\bibitem{Blanke:2012tv}
  M.~Blanke, B.~Shakya, P.~Tanedo and Y.~Tsai,
  JHEP {\bf 1208}, 038 (2012)
  [\href{http://arxiv.org/abs/arXiv:1203.6650}{arXiv:1203.6650  [hep-ph]}].

\bibitem{gherghetta}
T.~Gherghetta and A.~Pomarol,
Nucl.\ Phys.\ B {\bf 586}, 141 (2000)
[\href{http://arxiv.org/abs/hep-ph/0003129}{hep-ph/0003129}].

\bibitem{grossman-neubert}
Y.~Grossman and M.~Neubert,
Phys.\ Lett.\ B {\bf 474}, 361 (2000)
[\href{http://arxiv.org/abs/hep-ph/9912408}{hep-ph/9912408}].

\bibitem{Duling:2010lqa}
  B.~Duling,
PhD thesis:
 ``The custodially protected Randall-Sundrum model: Global features and distinct flavor signatures.''
 Technical University Munich, 2010.

\bibitem{neubertRS}
S.~Casagrande, F.~Goertz, U.~Haisch, M.~Neubert and T.~Pfoh,
  JHEP {\bf 0810}, 094 (2008)
  [\href{http://arxiv.org/abs/arXiv:0807.4937}{arXiv:0807.4937  [hep-ph]}];
  M.~Bauer, S.~Casagrande, U.~Haisch and M.~Neubert,
  JHEP {\bf 1009}, 017 (2010)
  [\href{http://arxiv.org/abs/arXiv:0912.1625}{arXiv:0912.1625   [hep-ph]}].

\bibitem{ALEPH:2005ab}
  S.~Schael {\it et al.}  [ALEPH and DELPHI and L3 and OPAL and SLD and LEP Electroweak Working Group and SLD Electroweak Group and SLD Heavy Flavour Group Collaborations],
  Phys.\ Rept.\  {\bf 427}, 257 (2006)
  [\href{http://arxiv.org/abs/hep-ex/0509008}{hep-ex/0509008}].

\bibitem{burdman}
  G.~Burdman,
  Phys.\ Rev.\ D {\bf 66}, 076003 (2002)
  [\href{http://arxiv.org/abs/hep-ph/0205329}{hep-ph/0205329}];
  Phys.\ Lett.\ B {\bf 590}, 86 (2004)
  [\href{http://arxiv.org/abs/hep-ph/0310144}{hep-ph/0310144}].

\bibitem{agashe}
  K.~Agashe, G.~Perez and A.~Soni,
  Phys.\ Rev.\ D {\bf 71}, 016002 (2005)
  [\href{http://arxiv.org/abs/hep-ph/0408134}{hep-ph/0408134}].

\bibitem{Fitzpatrick:2007sa}
A.~L.~Fitzpatrick, G.~Perez and L.~Randall,
Phys.\ Rev.\ Lett.\  {\bf 100}, 171604 (2008)
[\href{http://arxiv.org/abs/arXiv:0710.1869}{arXiv:0710.1869  [hep-ph]}].

\bibitem{Huber:2000ie}
  S.~J.~Huber and Q.~Shafi,
  Phys.\ Lett.\ B {\bf 498}, 256 (2001)
  [\href{http://arxiv.org/abs/hep-ph/0010195}{hep-ph/0010195}];
S.~J.~Huber,
  Nucl.\ Phys.\ B {\bf 666}, 269 (2003)
[\href{http://arxiv.org/abs/hep-ph/0303183}{hep-ph/0303183}].

\bibitem{Agashe:2004ay}
K.~Agashe, G.~Perez and A.~Soni,
Phys.\ Rev.\ Lett.\  {\bf 93}, 201804 (2004)
[\href{http://arxiv.org/abs/hep-ph/0406101}{hep-ph/0406101}].

\bibitem{Archer:2011bk}
P.~R.~Archer, S.~J.~Huber and S.~Jager,
JHEP {\bf 1112}, 101 (2011)
[\href{http://arxiv.org/abs/arXiv:1108.1433}{arXiv:1108.1433 [hep-ph]}].

\bibitem{Lees:2012tva}
J.~P.~Lees {\it et al.}  [BaBar Collaboration],
Phys.\ Rev.\ D {\bf 86}, 032012 (2012)
[\href{http://arxiv.org/abs/arXiv:1204.3933}{arXiv:1204.3933  [hep-ph]}].

\bibitem{Ball:2004rg}
  P.~Ball and R.~Zwicky,
  Phys.\ Rev.\ D {\bf 71}, 014015 (2005)
  [\href{http://arxiv.org/abs/hep-ph/0406232}{hep-ph/0406232}],  and
Phys.\ Rev.\ D {\bf 71}, 014029 (2005)
[\href{http://arxiv.org/abs/hep-ph/0412079}{hep-ph/0412079}].

\bibitem{Colangelo:1995jv}
P.~Colangelo, F.~De Fazio, P.~Santorelli and E.~Scrimieri,
Phys.\ Rev.\ D {\bf 53}, 3672 (1996)
[Erratum-ibid.\ D {\bf 57}, 3186 (1998)]
[\href{http://arxiv.org/abs/hep-ph/9510403}{hep-ph/9510403}].

\bibitem{Bouchard:2013eph}
  C.~Bouchard {\it et al.}  [HPQCD Collaboration],
  Phys.\ Rev.\ D {\bf 88}, no. 5, 054509 (2013)
  [Erratum-ibid.\ D {\bf 88}, no. 7, 079901 (2013)]
[\href{http://arxiv.org/abs/arXiv:1306.2384}{arXiv:1306.2384 [hep-lat}].

\bibitem{Horgan:2013hoa}
  R.~R.~Horgan, Z.~Liu, S.~Meinel and M.~Wingate,
  Phys.\ Rev.\ D {\bf 89}, 094501 (2014)
[\href{http://arxiv.org/abs/arXiv:1310.3722}{arXiv:1310.3722 [hep-lat}].

\bibitem{Feldmann:1999uf}
  T.~Feldmann, P.~Kroll and B.~Stech,
  Phys.\ Rev.\  D {\bf 58}, 114006 (1998)
   [\href{http://arxiv.org/abs/hep-ph/9802409}{hep-ph/9802409}]  and
  Phys.\ Lett.\  B {\bf 449}, 339 (1999)
     [\href{http://arxiv.org/abs/hep-ph/9812269}{hep-ph/9812269}];
  T.~Feldmann,
  Int.\ J.\ Mod.\ Phys.\  A {\bf 15}, 159 (2000)
    [\href{http://arxiv.org/abs/hep-ph/9907491}{hep-ph/9907491}].

\bibitem{DeFazio:2000my}
  F.~De Fazio and M.~R.~Pennington,
  JHEP {\bf 0007}, 051 (2000)
  [\href{http://arxiv.org/abs/hep-ph/0006007}{hep-ph/0006007}].

\bibitem{Ambrosino:2006gk}
 F.~Ambrosino {\it et al.}  [KLOE Collaboration],
 Phys.\ Lett.\ B {\bf 648}, 267 (2007)
[\href{http://arxiv.org/abs/hep-ex/0612029}{hep-ex/0612029}].
%
\bibitem{Ambrosino:2009sc}
  F.~Ambrosino {\it et al.},
  JHEP {\bf 0907}, 105 (2009)
    [\href{http://arxiv.org/abs/arXiv:0906.3819}{arXiv:0906.3819 [hep-ph]}].

\bibitem{Carlucci:2009gr}
A discussion of  this assumption can be found in
M.~V.~Carlucci, P.~Colangelo and F.~De Fazio,
  Phys.\ Rev.\ D {\bf 80}, 055023 (2009)
[\href{http://arxiv.org/abs/arXiv:0907.2160}{arXiv:0907.2160 [hep-ph]}].
%
\bibitem{Colangelo:2010bg}
  P.~Colangelo, F.~De Fazio and W.~Wang,
  Phys.\ Rev.\ D {\bf 81}, 074001 (2010)
  [\href{http://arxiv.org/abs/arXiv:1002.2880}{arXiv:1002.2880 [hep-ph]}].

\end{thebibliography}
\end{document}